# Second-order difference-frequency effects on floating wind turbine responses: full-QTF modelling and experimental comparison

Yi Zhang [a,b], Tom Tosdevin [c], Tianning Tang [a], Peter Stansby [a*]

[a] School of Engineering, The University of Manchester, Manchester M13 9PL, UK
[b] State Key Laboratory of Coastal and Offshore Engineering, Dalian University of Technology, Dalian 116024, China
[c] Department of Engineering Science, University of Oxford, Oxford OX1 2JD, UK

*Corresponding author: p.k.stansby@manchester.ac.uk (Peter Stansby)

## Abstract

Accurate prediction of low-frequency responses is important for the dynamic assessment of floating offshore wind turbines (FOWTs), as second-order difference-frequency wave loads can contribute substantially to platform motions and mooring loads. Their relative importance, however, remains less well established when aerodynamic and hydrodynamic low-frequency excitations coexist under operating conditions. This study develops an efficient time-domain formulation for full difference-frequency quadratic transfer function (QTF) loading using bandwidth truncation and integrates it into a fully coupled model of the IEA 15 MW turbine on the VolturnUS-S semi-submersible platform. Numerical predictions are compared with 1:70-scale real-time hybrid wave-basin experiments covering both irregular wave-only and combined turbulent wind–wave conditions. Three hydrodynamic formulations are compared: linear wave excitation, Newman's approximation, and the full-QTF model. Under wave-only conditions, second-order difference-frequency loading substantially improves the prediction of surge and the associated mooring response, particularly in energetic sea states, with the full-QTF model generally providing the closest agreement with measured low-frequency spectra and peak-amplitude distributions. Under combined wind–wave conditions, low-frequency thrust fluctuations dominate near rated wind speed, reducing the relative importance of difference-frequency wave excitation, whereas its contribution becomes more apparent at higher wind speeds as thrust fluctuations diminish. Overall, the full-QTF model provides the most consistent improvement over the linear and Newman models, with maximum surge standard deviation discrepancy from measurement under the most energetic wave conditions less than 18%, with smaller discrepancies for heave, pitch, and mooring tensions. The accuracy of efficient modelling is thus assessed for a typical FOWT including extreme waves.

*Keywords*: Floating offshore wind turbine; Difference-frequency loads; Quadratic transfer function; Low-frequency response; Wave-basin experiments

## 1. Introduction

Offshore wind is increasingly moving into deeper waters, where floating offshore wind turbines (FOWTs) provide access to extensive wind resources beyond the practical range of fixed-bottom foundations (Edwards et al., 2023). Unlike fixed-bottom turbines, FOWTs are compliant systems whose platform motions interact with aerodynamic loading, wave excitation, mooring restoring forces, and turbine control, giving rise to strongly coupled system dynamics. Reliable prediction of these coupled responses is therefore essential for their design and assessment. For semi-submersible FOWTs, the surge and pitch natural frequencies are typically designed to lie below the primary wave-frequency range (Wang et al., 2021), where first-order wave excitation alone may be insufficient to reproduce the observed resonant response. Second-order wave–body interactions involving pairs of incident-wave frequency components generate difference-frequency loads, including mean and slowly varying drift forces, which can excite these lightly damped low-frequency modes and significantly influence platform excursions and mooring loads (Coulling et al., 2013; Zhang et al., 2020). Their relative importance varies with platform characteristics and environmental conditions and can become particularly evident when aerodynamic excitation is weak or absent (Roald et al., 2013; Azcona et al., 2019).

Within second-order potential-flow theory, second-order wave-excitation loads are represented by quadratic transfer functions (QTFs), which characterise the sum- and difference-frequency load components arising from pairwise interactions between incident-wave components (Kim and Yue, 1990). Since the present study focuses on the low-frequency response of the floating system, only the difference-frequency QTFs are considered. Newman's approximation (Newman, 1974) provides an efficient alternative to the full QTF by estimating off-diagonal interactions from the diagonal mean-drift coefficients. Its accuracy depends on how well these diagonal coefficients represent the off-diagonal QTF over the wave-frequency pairs contributing to the low-frequency response of interest, which is generally favoured by a smoothly varying QTF in the relevant frequency region. Coulling et al. (2013) demonstrated the importance of including difference-frequency excitation when reproducing the global response of a semi-submersible FOWT. Subsequent experimental and numerical studies showed that Newman's approximation may underpredict slow-drift responses under certain conditions, whereas using the full difference-frequency QTF can improve predictions of low-frequency response (Simos et al., 2018; Yang et al., 2021). Since difference-frequency loads can excite low-frequency global platform motions, their representation can also affect the dynamic response of the mooring system and the resulting mooring-line tensions (Zhang et al., 2020; Li and Bachynski-Polić, 2021). These findings motivate a direct comparison of the full difference-frequency QTF and Newman's approximation for platform and mooring responses.

Accurate prediction of the low-frequency response depends not only on second-order wave excitation but also on the representation of damping and other hydrodynamic effects near the platform natural frequencies. Previous studies have shown that discrepancies between numerical predictions and experimental measurements of low-frequency platform motions can persist, owing in part to uncertainties in viscous drag and damping and to other hydrodynamic effects not fully captured by conventional potential-flow models (Pegalajar-Jurado and Bredmose, 2019; Robertson et al., 2020; Li and Bachynski-Polić, 2021; Wang et al., 2022). More recently, a comparative study involving a wide range of numerical approaches, including high-fidelity Navier–Stokes solvers, examined the hydrodynamic response of a FOWT and reported substantial differences in the predicted surge and pitch responses and mooring loads (Yu et al., 2025). Full-QTF modelling should therefore be regarded as an improved representation of second-order potential-flow excitation rather than a complete description of low-frequency hydrodynamics. This raises the question of how much discrepancy remains between numerical predictions and experiments once both first- and second-order hydrodynamic loads are included, especially under energetic wave conditions. Relatively few studies have quantified these residual discrepancies against experimental measurements for the coupled platform and mooring responses of large semi-submersible FOWTs.

Assessing the relative benefit of the full QTF becomes more complex under operating conditions, where aerodynamic loading also influences the low-frequency dynamics of the floating system. Turbulent rotor-thrust fluctuations can directly excite low-frequency surge and pitch, while aerodynamic damping and turbine control modify the corresponding resonant response. Duarte et al. (2014) numerically investigated the effects of second-order hydrodynamic loading in conjunction with aerodynamic excitation under different wind–wave conditions. Using software-in-the-loop hybrid experiments, Azcona et al. (2019) further showed that strong turbulent-thrust fluctuations can mask the contribution of second-order hydrodynamic excitation to the low-frequency platform response, whereas this contribution becomes more apparent when aerodynamic excitation is reduced. More recent studies have further investigated the effects of aerodynamic and second-order hydrodynamic loading under combined wind–wave conditions, showing that their relative importance can vary with the floating configuration and environmental condition (Mahfouz et al., 2021; Shi et al., 2023). However, the relative performance of Newman's approximation and the full difference-frequency QTF under turbulent aerodynamic excitation has received less attention, with relatively limited experimental evidence available for operating FOWTs under turbulent wind–wave conditions.

A further practical challenge is the computational cost of evaluating full-QTF loads in long, fully coupled time-domain simulations, as the conventional formulation involves summation over all relevant wave-frequency pairs and therefore becomes increasingly expensive as the number of frequency components increases. Duarte et al. (2014) reduced this cost by grouping QTF contributions associated with the same difference frequency and reconstructing the resulting second-order force time histories through an inverse FFT before the coupled FAST simulation. Alternative approaches have exploited the narrow-band nature of slow-drift resonance by retaining only selected regions of the QTF matrix associated with response-relevant difference frequencies (Simos et al., 2018). Motivated by this principle, the present study adopts a difference-frequency bandwidth truncation within the full-QTF time-domain formulation. Only frequency-pair interactions satisfying a prescribed difference-frequency limit are retained, thereby reducing the number of QTF terms while preserving those associated with the low-frequency response of interest. A low-rank representation of the QTF matrix is also briefly considered as a possible route to reducing the computational cost, but it is not used in the coupled simulations presented here.

Coupled time-domain tools such as OpenFAST, SIMA, and OrcaFlex are widely used for FOWT analysis and include capabilities for modelling second-order difference-frequency wave loads, e.g., Otter et al. (2022). The present work builds on the previously developed in-house OREGEN coupled modelling framework (Zhang et al., 2025, 2026a), which integrates aerodynamic loading, rotor control, hydrodynamics, platform dynamics, and mooring dynamics in an efficient explicit time-domain formulation. Its hydrodynamic and platform-response predictions have been validated against wave-basin experiments involving focussed steep-wave conditions in the absence of wind loading (Zhang et al., 2025), while the aerodynamic and rotor-control models have been validated under steady operating conditions and implemented for simulations under irregular-wave and turbulent-wind conditions (Zhang et al., 2026a). A second-order difference-frequency excitation model is newly integrated into this framework and evaluated for the IEA 15 MW reference wind turbine (Gaertner et al., 2020) supported by the VolturnUS-S semi-submersible platform (Allen et al., 2020). In this paper, the extended framework is validated against 1:70-scale wave-basin experiments covering both wave-only and combined turbulent wind–wave conditions (Zhang et al., 2026b). For the combined cases, real-time hybrid testing coupled the physical platform and mooring system with numerical aerodynamic and rotor-control models, with measured platform motions fed back in real time and the resulting rotor thrust reproduced at hub height by a pair of controlled drone propellers.

Three hydrodynamic formulations are compared: linear wave excitation (extended to include time-dependent surge position), Newman's approximation, and the full difference-frequency QTF model. Their predictions are assessed against experimental measurements of platform motions and mooring tensions. The aims of this paper are (1) to determine the significance of second-order difference-frequency excitation for the platform and mooring responses under wave-only and turbulent wind–wave conditions, (2) to quantify the additional predictive benefit of retaining the full off-diagonal QTF structure relative to Newman's approximation and assess how this benefit changes in the presence of low-frequency aerodynamic excitation, and (3) to quantify the remaining discrepancies between numerical predictions and experiments under energetic wave conditions when both first- and second-order hydrodynamic loads are included.

The remainder of this paper is organised as follows. Section 2 presents the second-order difference-frequency force formulations, including the full-QTF model, bandwidth truncation, low-rank acceleration, and Newman's approximation. Section 3 describes the experimental and numerical models and the test campaign. Section 4 presents the numerical–experimental comparisons for wave-only and combined wind–wave conditions and examines the effect of simulation duration. Section 5 summarises the main findings and conclusions of the study.

## 2. Calculation of second-order difference-frequency forces

### 2.1. Full-QTF formulation

The free-surface elevation of a unidirectional irregular wave at the reference location is represented as the superposition of *N* linear wave components,

$$\eta(t) = \mathrm{Re}\{\textstyle\sum_{i=1}^{N} z_i(t)\} \tag{1}$$

where the *i*-th wave component is expressed in complex form as:

$$z_i(t) = a_i \exp[\mathrm{i}\theta_i(t)] \tag{2}$$

where $a_i = \sqrt{2S(\omega_i)\delta\omega}$ is the wave amplitude, determined from the discretised wave spectrum; $\theta_i(t) = -\omega_i t + \phi_i$ is the phase angle; $t$ is time; $\omega_i$ and $\phi_i$ are the angular frequency and initial random phase, respectively, and $\mathrm{i} = \sqrt{-1}$ is imaginary unit. The complex wave-component vector is then defined as:

$$\mathbf{z}(t) = [z_1(t), z_2(t), \dots, z_N(t)]^{\mathrm{T}} \tag{3}$$

where the superscript $T$ denotes the transpose.

Second-order wave loads include both sum- and difference-frequency contributions. This study focuses on the latter due to their importance in low-frequency platform and mooring responses. For a difference-frequency QTF, $Q_{ij}$ denotes the complex transfer coefficient associated with the interaction between the *i*-th and *j*-th wave components at the difference frequency $\omega_i - \omega_j$ (Pinkster, 1980). It can be expressed as:

$$Q_{ij} = |Q_{ij}| \exp(\mathrm{i}\psi_{ij}) \tag{4}$$

where $|Q_{ij}|$ and $\psi_{ij}$ are its magnitude and phase, respectively. The complete QTF matrix $Q$ is Hermitian, i.e.,

$$\mathbf{Q} = \mathbf{Q}^{\mathrm{H}};\ Q_{ji} = Q_{ij}^{*} \tag{5}$$

where the superscript $H$ denotes the conjugate transpose and $*$ denotes the complex conjugate. Consequently, the diagonal entries are real-valued, i.e., $\mathrm{Im}(Q_{ii}) = 0$.

For an arbitrary generalised force component (force or moment), the second-order difference-frequency contribution can be expressed in matrix form as:

$$F^{(2-)}(t) = \mathbf{z}^{\mathrm{T}}(t)\,\mathbf{Q}\,\mathbf{z}^{*}(t) \tag{6}$$

The corresponding component-wise form is:

$$F^{(2-)}(t) = \sum_{i=1}^{N}\sum_{j=1}^{N} Q_{ij}\, z_i(t) z_j^{*}(t) \tag{7}$$

The complex contribution associated with the $(i, j)$ frequency pair is:

$$Q_{ij} z_i(t) z_j^{*}(t) = a_i a_j \left|Q_{ij}\right| \exp\left\{\mathrm{i}\left[\theta_i(t) - \theta_j(t) + \psi_{ij}\right]\right\} \tag{8}$$

Owing to the Hermitian symmetry of $Q$, i.e., $Q_{ji} = Q_{ij}^{*}$, the contributions from the $(i, j)$ and $(j, i)$ frequency pairs are complex conjugates, and their sum is therefore twice the real part of either contribution. Consequently, the total force in Eq. (7) is real-valued, and only the diagonal entries and the lower triangular part $(i > j)$ of the QTF matrix need to be evaluated:

$$F^{(2-)}(t) = \sum_{i=1}^{N} Q_{ii}\, |z_i(t)|^2 + 2\mathrm{Re}\left\{\sum_{i=2}^{N}\sum_{j=1}^{i-1} Q_{ij} z_i(t) z_j^{*}(t)\right\} \tag{9}$$

The diagonal terms correspond to the zero-frequency mean-drift contribution, whereas the off-diagonal terms generate the oscillatory difference-frequency contributions.

Although the Hermitian symmetry of the QTF reduces the number of independent frequency-pair evaluations from $N^2$ to $N(N+1)/2$, the computational complexity remains $\mathcal{O}(N^2)$ per time step, or $\mathcal{O}(N_t N^2)$ over $N_t$ time steps. For finely resolved wave representations, $N$ can be of the order of $10^3$, making the per-step evaluation computationally expensive, particularly when repeated over long-duration simulations with large $N_t$. To further reduce the computational cost of full-QTF time-domain evaluation, two acceleration strategies are discussed below: difference-frequency bandwidth truncation and low-rank approximation. Both approaches are briefly described, while only the former is adopted in the present simulations. The detailed implementation and performance assessment of the latter are presented in a separate study.

### *2.1.1. Difference-frequency bandwidth truncation*

The second-order difference-frequency response of a floating system is primarily governed by interactions between wave components whose frequency differences fall within the low-frequency range relevant to the platform dynamics, owing to the relatively low natural frequencies of floating platforms. In the QTF matrix, these low difference-frequency interactions correspond to entries near the main diagonal, whereas entries farther from the diagonal are associated with larger difference frequencies. Therefore, restricting the evaluated frequency pairs to a prescribed difference-frequency bandwidth can substantially reduce the number of QTF evaluations while retaining the interactions most relevant to the low-frequency response. Accordingly, the bandwidth-truncated difference-frequency force is evaluated as:

$$F_{\mathrm{band}}^{(2-)}(t) = \sum_{i=1}^{N} Q_{ii} \, |z_i(t)|^2 + 2 \,\mathrm{Re} \left\{ \sum_{i=2}^{N} \sum_{\substack{j=1 \\ |\omega_i - \omega_j| \le \Omega_{\mathrm{cut}}}}^{i-1} Q_{ij} \, z_i(t) z_j^*(t) \right\} \tag{10}$$

Here, $\Omega_{\mathrm{cut}}$ denotes the maximum retained angular difference frequency, related to the difference-frequency bandwidth $\Delta f$ by $\Omega_{\mathrm{cut}} = 2\pi\Delta f$. The diagonal terms are always retained because they correspond to the zero-frequency mean-drift contribution, while only off-diagonal interactions satisfying $|\omega_i - \omega_j| \le \Omega_{\mathrm{cut}}$ are included. The computational complexity is reduced from $\mathcal{O}(N^2)$ to $\mathcal{O}(NM)$, where $M = \lfloor \Omega_{\mathrm{cut}}/\delta\omega \rfloor = \lfloor \Delta f/\delta f \rfloor$, with $\delta\omega = 2\pi\delta f$ denoting the angular-frequency resolution of the wave discretisation. The bandwidth $\Delta f$ is determined through a convergence study by progressively increasing the retained bandwidth until further increases produce negligible changes in the quantities of interest, thereby balancing computational efficiency and the accuracy of the resulting second-order response.

### *2.1.2. Low-rank acceleration*

Since the QTF matrix $Q$ is Hermitian, it admits an eigendecomposition. When its eigenvalue spectrum decays rapidly, a rank-$R$ approximation can be constructed by retaining the $R$ dominant modes:

$$\mathbf{Q} \approx \mathbf{Q}_R = \mathbf{U}_R \mathbf{\Lambda}_R \mathbf{U}_R^{\mathrm{H}} \tag{11}$$

The eigenvalues are ordered by decreasing magnitude,

$$|\lambda_1| \ge |\lambda_2| \ge \cdots \ge |\lambda_N| \tag{12}$$

Here, $R \ll N$ when the eigenvalue spectrum decays sufficiently rapidly, $\mathbf{U}_R = [\mathbf{u}_1, \mathbf{u}_2, \ldots, \mathbf{u}_R]$ contains the eigenvectors associated with the $R$ dominant eigenvalues, and $\mathbf{\Lambda}_R = \mathrm{diag}(\lambda_1, \ldots, \lambda_R)$ contains the corresponding real eigenvalues. The truncation rank $R$ is selected as the minimum number of dominant modes required to satisfy a prescribed cumulative squared-eigenvalue ratio:

$$\frac{\sum_{r=1}^{R} |\lambda_r|^2}{\sum_{r=1}^{N} |\lambda_r|^2} \ge \tau \tag{13}$$

where $\tau$ denotes the prescribed retention threshold, typically chosen close to unity (e.g., 0.99 or 0.999).

Substituting the truncated eigen decomposition in Eq. (11) into the full-QTF formulation in Eq. (6) gives:

$$F^{(2-)}(t) \approx \mathbf{z}^{\mathrm{T}}(t) \mathbf{U}_R \mathbf{\Lambda}_R \mathbf{U}_R^{\mathrm{H}} \mathbf{z}^*(t) \tag{14}$$

Define the complex projection:

$$\mathbf{p}(t) = \mathbf{U}_R^{\mathrm{T}} \mathbf{z}(t) \tag{15}$$

Then Eq. (14) can be written as:

$$F^{(2-)}(t) \approx \mathbf{p}^{\mathrm{T}}(t)\mathbf{\Lambda}_R\mathbf{p}^*(t) = \sum_{r=1}^{R} \lambda_r \, |p_r(t)|^2 \tag{16}$$

with

$$p_r(t) = \mathbf{u}_r^{\mathrm{T}}\mathbf{z}(t) = \sum_{i=1}^{N} u_{ir} z_i(t) \tag{17}$$

where $u_{ir}$ denotes the $i$-th component of the $r$-th eigenvector, $\mathbf{u}_r = [u_{1r}, u_{2r}, \dots, u_{Nr}]^{\mathrm{T}}$.

A comparison of Eqs. (16) and (9) indicates that the computational complexity is reduced from $\mathcal{O}(N^2)$ to $\mathcal{O}(NR)$. When the QTF eigenvalue spectrum decays rapidly, only a limited number of dominant modes are required, allowing the time-domain evaluation to be substantially accelerated.

2.2. Newman's approximation

Newman (1974) proposed an approximation for slowly varying second-order forces based solely on the diagonal QTFs, assuming that the off-diagonal QTF varies smoothly near the main diagonal. Since the QTF matrix is Hermitian, its diagonal entries $Q_{ii}$ are real-valued. Following Standing et al. (1987), the off-diagonal QTFs are estimated from the corresponding diagonal values using the signed geometric-mean approximation, i.e.

$$Q_{ij}^{\mathrm{N}} = \frac{1}{2}\left[\mathrm{sgn}(Q_{ii}) + \mathrm{sgn}(Q_{jj})\right]\sqrt{|Q_{ii}Q_{jj}|} \tag{18}$$

where $\mathrm{sgn}(\cdot)$ denotes the sign function, taking values 1, 0, and $-1$ for positive, zero, and negative arguments, respectively. Accordingly, interactions associated with diagonal QTFs of opposite signs vanish, whereas same-sign pairs contribute according to their common sign. Substituting Eq. (18) into the full-QTF formulation in Eq. (7) and separating the remaining contributions according to the sign of the diagonal QTFs gives:

$$F_{\mathrm{N}}^{(2-)}(t) = \sum_{\substack{i=1 \\ Q_{ii}\geq 0}}^{N} \sum_{\substack{j=1 \\ Q_{jj}\geq 0}}^{N} \left[\sqrt{Q_{ii}} z_i(t)\right]\left[\sqrt{Q_{jj}} z_j(t)\right]^* - \sum_{\substack{i=1 \\ Q_{ii}<0}}^{N} \sum_{\substack{j=1 \\ Q_{jj}<0}}^{N} \left[\sqrt{-Q_{ii}} z_i(t)\right]\left[\sqrt{-Q_{jj}} z_j(t)\right]^* \tag{19}$$

Since each summand is separable in $i$ and $j$, Eq. (19) can be factorised as:

$$F_{\mathrm{N}}^{(2-)}(t) = \left|\sum_{\substack{i=1 \\ Q_{ii}\geq 0}}^{N} \sqrt{Q_{ii}}\, z_i(t)\right|^2 - \left|\sum_{\substack{i=1 \\ Q_{ii}<0}}^{N} \sqrt{-Q_{ii}}\, z_i(t)\right|^2 \tag{20}$$

Thus, the original double summation is replaced by two single summations, reducing the computational complexity from $\mathcal{O}(N^2)$ to $\mathcal{O}(N)$. However, this efficiency is achieved at the expense of the full off-diagonal QTF information, including its phase variation, and the approximation may become less accurate when the QTF varies rapidly in the vicinity of the main diagonal or exhibits significant off-diagonal structure.

## 3. Floating wind turbine model

### 3.1. *Experimental scale model*

A 1:70 Froude-scaled model of the IEA 15 MW reference wind turbine (Gaertner et al., 2020) mounted on the VolturnUS-S semi-submersible platform (Allen et al., 2020) was tested in the COAST Laboratory Ocean Basin at the University of Plymouth. Fig. 1 shows the schematic of the 15 MW reference FOWT and the experimental setup of the 1:70-scale model in the basin. The basin is 35 m long and 15.65 m wide, with a water depth of 2.86 m in the working area, corresponding to the 200 m design water depth of the full-scale system. The platform was restrained by three catenary mooring lines, with the mooring lengths truncated to accommodate the basin dimensions while retaining the principal low-frequency restoring characteristics of the reference system. Tables 1 and 2 summarise the key properties of the FOWT model and the mooring system, respectively. The moments of inertia and centre of mass (CoM) position are taken from Ransley et al. (2022), who reported the same FOWT model used in the present experiments. Both model-scale values and their corresponding full-scale prototype values are provided to facilitate consistent comparison between the numerical and experimental models.

To avoid the Reynolds-number scaling limitations associated with a geometrically scaled rotor, a software-in-the-loop (SIL) approach (e.g., Azcona et al., 2019; Bonnefoy et al., 2024) was employed. The rotor thrust was calculated numerically in real time using the in-house aero-servo solver OREGEN_BEMT (Zhang et al., 2026a), with the prescribed turbulent wind field and measured platform motions as inputs. The resulting thrust was scaled to model scale and reproduced at hub height by a thruster comprising two drone propellers (Fig. 1(b)). Prior to the experiments, the thruster was calibrated to establish the relationship between the applied thrust force and the pulse width modulation (PWM) signal. Platform six-degree-of-freedom motions were measured using a Qualisys optical motion-capture system. Mooring-line tensions were measured by submersible load cells installed at the fairleads, while the incident wave field was monitored using resistance-type wave gauges. Details of the scale model, mooring configuration, experimental setup, and real-time aerodynamic thrust emulation are provided in Zhang et al. (2026b).

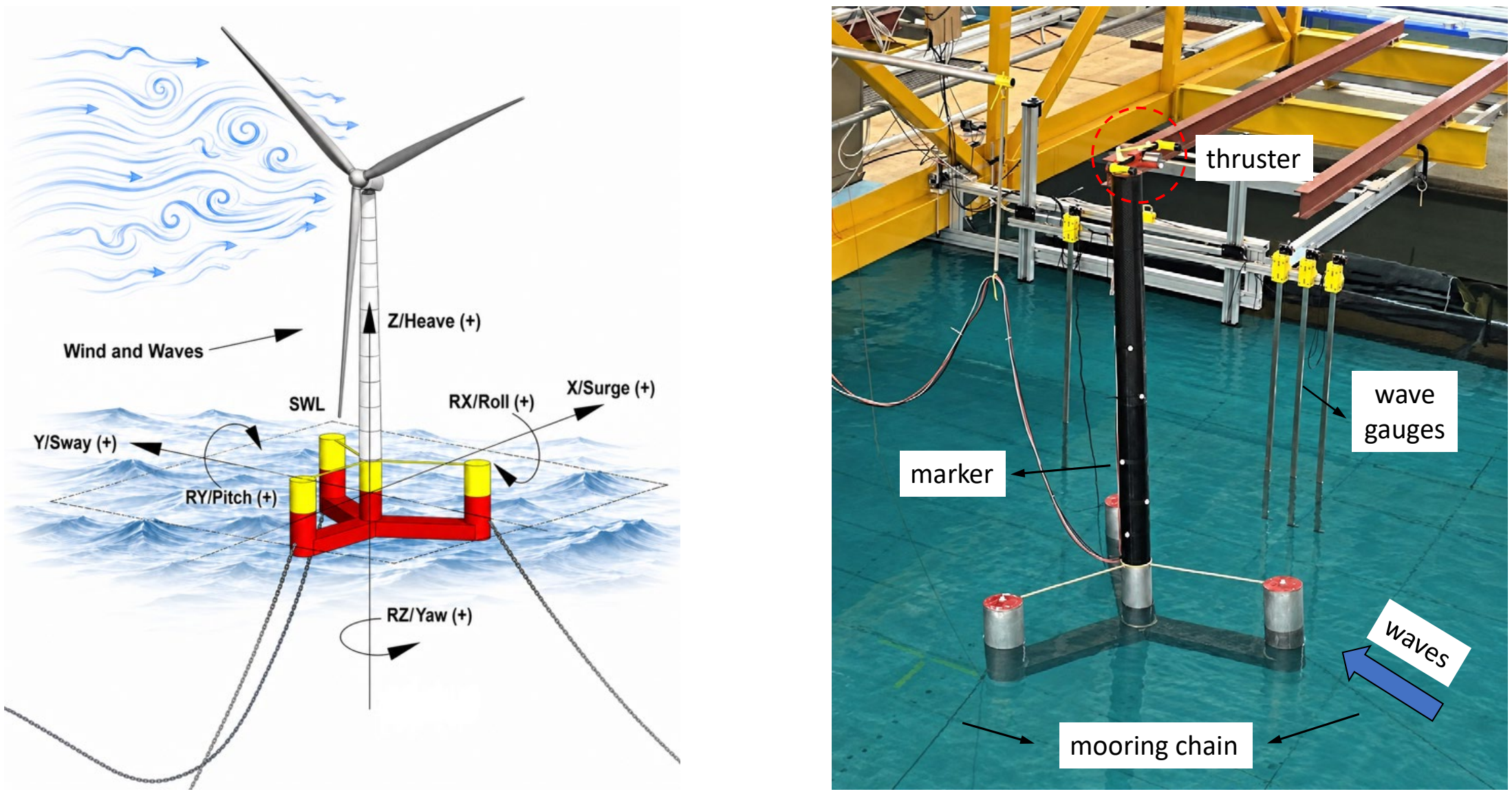

(a) (b)

**Fig. 1.** (a) Schematic of the 15 MW reference FOWT, adapted from Allen et al. (2020). (b) Experimental setup of the 1:70-scale FOWT model in the COAST Ocean Basin, adapted from Zhang et al. (2026b).

**Table 1.** Key properties of the FOWT model at 1:70 model scale and full scale.

| Parameter | Model scale | Full scale |
|---|---|---|
| Total system mass | 56.3 kg | 1.9311e7 kg |
| Roll / pitch inertia about CoM | 26.68 kg·m$^2$ | 4.4841e10 kg·m$^2$ |
| CoM position from SWL | −0.022 m | −1.54 m |
| Draft | 0.286 m | 20 m |

**Table 2.** Mooring system properties and configuration. Note that ($x$, $y$, $z$) coordinates are with respect to the centre of the FOWT model in the horizontal plane and the SWL in the vertical direction.

| Parameter | Model scale | Full scale |
|---|---|---|
| Dry line density | 0.144 kg/m | 705.6 kg/m |
| Fore mooring length | 9.685 m | 677.95 m |
| Port- and starboard-aft mooring length | 9.017 m | 631.19 m |
| Volume-equivalent diameter | 4.84 mm | 0.339 m |
| Fore fairlead position ($x$, $y$, $z$) | (−0.839, 0.0, −0.2) m | (−58.744, 0.0, −14.0) m |
| Port-aft fairlead position ($x$, $y$, $z$) | (0.420, −0.727, −0.2) m | (29.372, −50.876, −14.0) m |
| Starboard-aft fairlead position ($x$, $y$, $z$) | (0.420, 0.727, −0.2) m | (29.372, 50.876, −14.0) m |
| Fore anchor position ($x$, $y$, $z$) | (−9.595, 0.0, −2.86) m | (−671.65, 0.0, −200.0) m |
| Port-aft anchor position ($x$, $y$, $z$) | (4.342, −7.655, −2.86) m | (303.94, −534.45, −200.0) m |
| Starboard-aft anchor position ($x$, $y$, $z$) | (4.342, 7.655, −2.86) m | (303.94, 534.45, −200.0) m |
| Extensional stiffness ($EA$) | 9.534 kN | 3270 MN |

### *3.2. Numerical modelling framework*

#### *3.2.1. Coupled numerical model*

Building on our previously developed in-house OREGEN modelling framework for FOWTs (Zhang et al., 2025, 2026a), this study extends the model to incorporate second-order difference-frequency wave loads following the methodology presented in Section 2. As shown in Fig. 2, the coupled numerical model integrates wind inflow, aerodynamics, rotor dynamics, rotor control, hydrodynamics, platform dynamics, and mooring dynamics within a unified time-domain framework, with the coupled interactions resolved through the exchange of loads, platform states, and control signals among the relevant subsystems at each time step. The aerodynamic and control modules determine the rotor thrust and torque, while the hydrodynamic module provides the first-order wave loads and second-order difference-frequency wave loads. These loads, together with the mooring loads, drive the platform response, which is fed back to the aerodynamic, control, and mooring subsystems to account for the fully coupled system dynamics.

A numerical model was developed to represent the floating wind turbine tested in the wave basin, with the platform properties, mooring configuration, and environmental conditions corresponding to their full-scale equivalents based on Froude scaling. The platform's linear frequency-domain hydrodynamic coefficients were computed using the in-house boundary element method (BEM) solver OREGEN-BEM (Li and Stansby, 2023), previously validated against WAMIT results in Zhang et al. (2025). The wave excitation coefficients were used to reconstruct the first-order wave excitation force in the time domain, while the added mass and radiation damping coefficients were incorporated through the Cummins formulation to account for radiation effects. The first-order wave excitation forces were evaluated at the time-varying platform surge position to account for displacement-dependent changes in the incident-wave loading. This treatment has been shown to improve motion prediction in focussed waves (Zhang et al., 2025) and is also found here to improve predictions in irregular waves. Second-order difference-frequency loads were introduced separately using either Newman's approximation or the difference-frequency QTFs, as given in Eqs. (20) and (10), respectively. The motion-induced viscous damping was represented through empirical quadratic Morison-type drag terms based on the platform motion velocity, with the total viscous drag force obtained by summing the contributions from individual platform sections. Details of the formulation and the selection of the drag coefficients were provided in Zhang et al. (2025). A numerical model of the mooring system was developed in MoorDyn (Hall and Goupee, 2015) based on the mooring properties and configuration given in Table 2. Each mooring line was discretised into 40 segments to provide sufficient spatial resolution of the line dynamics, with further refinement producing negligible changes in the predicted responses. Following the MoorDyn settings in the reference OpenFAST model for this FOWT (Barter et al., 2026), the normal and tangential drag coefficients of the lines were set to 2.0 and 0.4, respectively, while the corresponding added mass coefficients were set to 0.8 and 0.25 to account for the hydrodynamic forces on the mooring lines. The aerodynamics of the IEA 15 MW reference wind turbine were modelled using OREGEN_BEMT, an in-house blade-element momentum theory (BEMT) solver based on the blade geometry and airfoil aerodynamic data reported by Gaertner et al. (2020). To ensure consistency with the wave-basin tests, the same aerodynamic model, rotor controller, and wind files used in the SIL computations were employed in the numerical simulations. The blades and tower were assumed rigid. A reference-tracking proportional–integral (PI) controller with platform-velocity feedback was adopted to regulate the blade pitch and generator torque. The platform-velocity feedback with gain scheduling was introduced to mitigate negative aerodynamic damping, suppress platform pitch motion near its natural frequency, and reduce rotor-speed fluctuations, as demonstrated numerically and experimentally in Zhang et al. (2026a, 2026b). Further details of the controller design and gain tuning are provided in Zhang et al. (2026a).

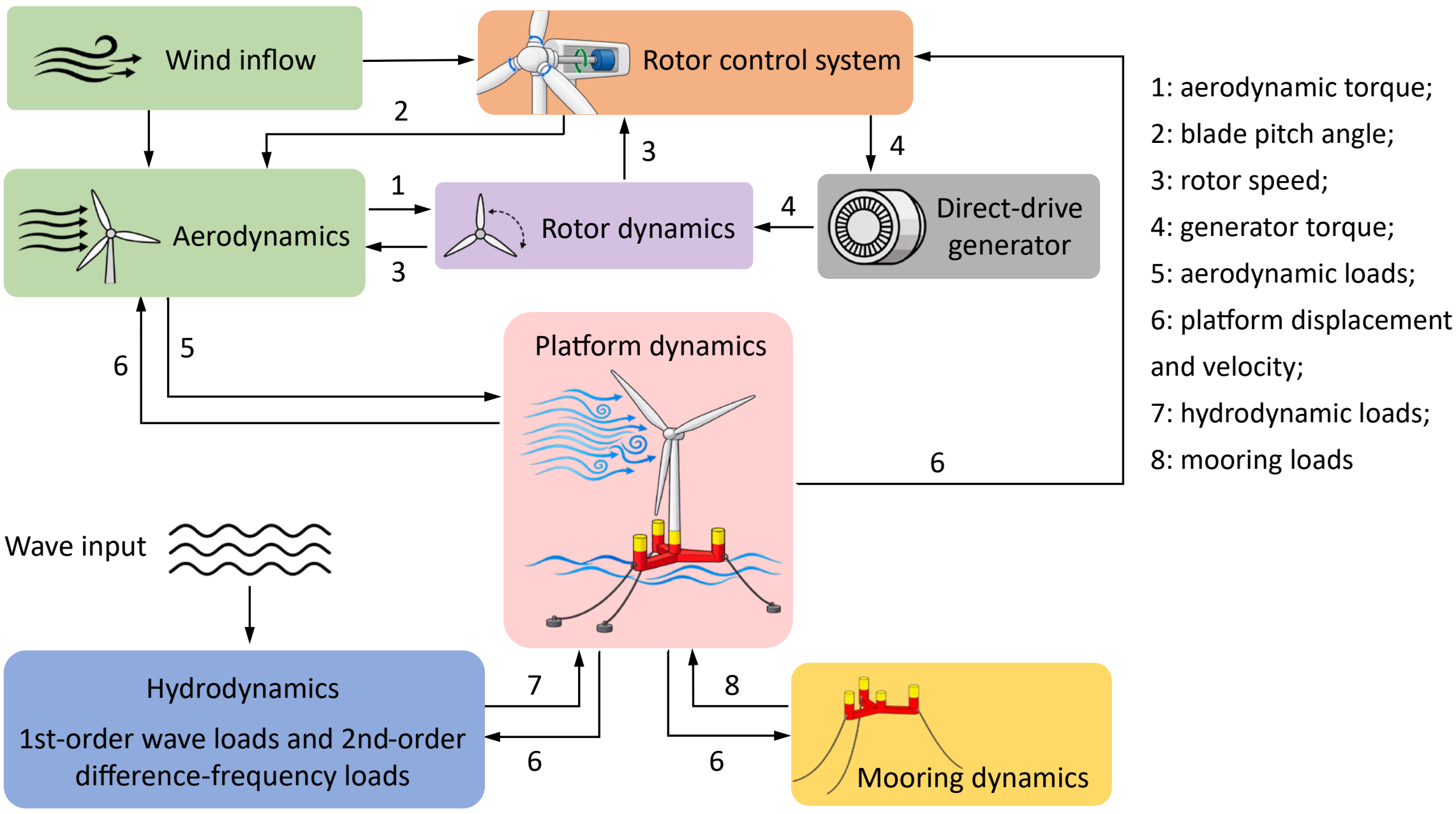

**Fig. 2.** Schematic diagram of the coupled numerical model, with the FOWT adapted from Allen et al. (2020).

### *3.2.2. Difference-frequency QTFs*

The difference-frequency QTFs, evaluated using WAMIT on a $46 \times 46$ frequency grid spanning approximately $0.04-0.40$ Hz, were obtained from the publicly available IEA-15-240-RWT repository

([https://github.com/IEAWindSystems/IEA-15-240-RWT/tree/master/OpenFAST/IEA-15-240-RWT-UMaineSemi/HydroData](https://github.com/IEAWindSystems/IEA-15-240-RWT/tree/master/OpenFAST/IEA-15-240-RWT-UMaineSemi/HydroData)). To obtain a finer frequency resolution consistent with the wave-frequency discretisation used in the time-domain simulations, bilinear interpolation was applied to the off-diagonal QTFs, while the mean-drift QTFs along the main diagonal were interpolated separately using one-dimensional linear interpolation. Fig. 3 shows the real and imaginary parts of the difference-frequency QTFs for surge, heave, and pitch modes. Note that the original QTF values are visualised using a symmetric inverse-hyperbolic-sine colour scale to enhance the visibility of small-magnitude values near the main diagonal. The solid black line marks the main diagonal, corresponding to the mean-drift QTFs, which form the basis of Newman's approximation, while the dashed black lines indicate the 0.075 Hz difference-frequency bandwidth used to limit off-diagonal QTF contributions and reduce computational cost. As expected from Hermitian symmetry, the real and imaginary parts of the difference-frequency QTFs are symmetric and antisymmetric about the main diagonal, respectively, with the imaginary component vanishing along the diagonal. The QTFs vary relatively smoothly near the diagonal, whereas more pronounced variations in the real and imaginary components are observed with increasing frequency separation, suggesting a stronger dependence of the complex QTF on the interacting frequency pair. The frequency dependence differs among surge, heave, and pitch, reflecting their different sensitivities to second-order wave interactions.

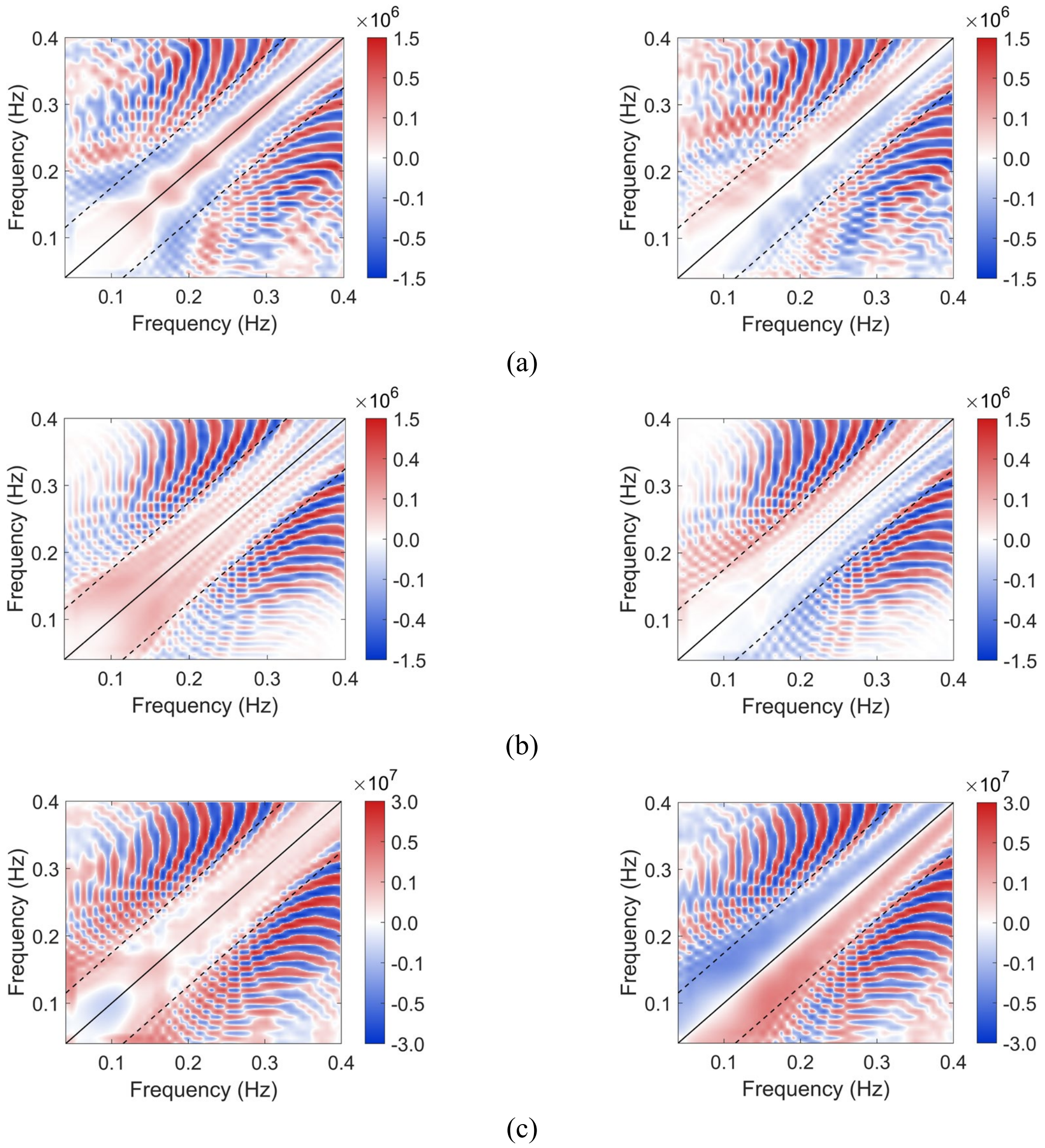

**Fig. 3.** Second-order difference-frequency QTFs for (a) surge, (b) heave, and (c) pitch. The real and imaginary parts are shown in the left and right panels, respectively. The colorbar units are N/m$^2$ for surge and heave, and N-m/m$^2$ for pitch. The solid black line denotes the main diagonal (mean-drift QTF), while the dashed black lines indicate a difference-frequency bandwidth of 0.075 Hz.

### *3.3. Test campaign*

Both wave-only and combined wind–wave cases are considered to investigate the effects of second-order difference-frequency wave forces. Irregular waves and turbulent winds are employed to represent realistic offshore environmental conditions. The wave-only cases isolate the wave-induced second-order difference-frequency response, whereas the combined cases assess the influence of low-frequency aerodynamic excitation from turbulent winds, which may overlap with the difference-frequency range and jointly affect the platform dynamics.

Table 3 lists the full-scale wave conditions for representative wave-only cases, including the significant wave height ($H_s$), peak wave period ($T_p$), peak enhancement factor ($\gamma$), and spectral wave steepness ($0.5k_pH_s$), where $k_p$ is the wavenumber corresponding to the spectral peak period $T_p$. The latter is included as an indicator of incident wave nonlinearity. The irregular waves were generated by discretising the target JONSWAP spectrum into a finite number of frequency components, with the amplitude of each component determined from the spectral density and its phase assigned randomly. In the experiments, the generated waves were calibrated to ensure good agreement between the measured and target wave spectra. The experimental and numerical irregular waves were therefore characterised by the same target spectral parameters, providing a consistent basis for comparing the statistical and spectral characteristics of the resulting platform motions and mooring-line tensions. Notably, Wave 5 represents the severe sea state under the parked condition. Table 4 presents the environmental conditions for the combined wind–wave DLC 1.6 cases, including the mean hub-height wind speed ($\bar{u}_{\mathrm{hub}}$), turbulence intensity (TI), and the corresponding wave parameters. The full-scale turbulent wind fields used in both the experiments and numerical simulations were generated using TurbSim (Jonkman, 2014) on a $21 \times 21$ grid for the specified $\bar{u}_{\mathrm{hub}}$ and TI. The environmental conditions considered in this study are derived from site-specific ERA5 reanalysis metocean data for the selected deployment site off the Gulf of Maine (Zhang et al., 2026b), with the analysed cases representing 50-year return-period conditions and critical extreme load scenarios for FOWT design and safety assessment.

**Table 3.** Full-scale wave conditions for different wave-only cases.

| Case ID | $H_s$ (m) | $T_p$ (s) | $\gamma$ | $0.5k_pH_s$ |
|---|---|---|---|---|
| Wave 1 | 1.4 | 7.0 | 1.0 | 0.058 |
| Wave 2 | 5.1 | 9.4 | 2.7 | 0.116 |
| Wave 3 | 6.3 | 12.0 | 1.3 | 0.088 |
| Wave 4 | 5.8 | 15.0 | 1.0 | 0.052 |
| Wave 5 | 9.0 | 13.5 | 1.8 | 0.099 |

**Table 4.** Full-scale wind and wave conditions for different combined cases.

| Case ID | $\bar{u}_{\mathrm{hub}}$ (m/s) | TI (%) | $H_s$ (m) | $T_p$ (s) | $\gamma$ | $0.5k_pH_s$ |
|---|---|---|---|---|---|---|
| DLC1.6_U8_2 | 8 | 17.4 | 6.3 | 12.0 | 1.3 | 0.088 |
| DLC1.6_U11_1 | 11 | 15.1 | 5.5 | 9.0 | 3.8 | 0.137 |
| DLC1.6_U11_2 | 11 | 15.1 | 6.8 | 12.2 | 1.4 | 0.092 |
| DLC1.6_U14_2 | 14 | 13.8 | 7.4 | 12.5 | 1.6 | 0.095 |
| DLC1.6_U20_1 | 20 | 12.4 | 6.6 | 10.0 | 3.6 | 0.133 |
| DLC1.6_U20_2 | 20 | 12.4 | 8.3 | 13.1 | 1.7 | 0.097 |

## 4. Numerical–experimental comparison and discussion

This section compares the experimental measurements with the numerical simulation results for the corresponding cases, ranging from wave-only to combined wind–wave conditions. Three numerical models are considered: the linear model, the model incorporating second-order difference-frequency loads using Newman's approximation, and the model using the full difference-frequency QTFs (Diff-QTF). All numerical simulations were performed at full scale, and the experimental results were converted to full-scale values for comparison. Each simulation was run for 3800 s, covering approximately 1 h of full-scale experimental data, with a time step of 0.05 s. For the Diff-QTF model, a difference-frequency bandwidth of 0.075 Hz was adopted to retain the dominant low-frequency second-order wave load components that contribute most strongly to the platform responses while reducing computational cost. The sensitivity analysis in Appendix A shows that further increasing the bandwidth has only a limited influence on the predicted responses, while the time-step convergence study is provided in Appendix B.

With the numerical settings adopted in the present study, a typical 3800-s combined wind–wave simulation required approximately 30 min of wall-clock time on a laptop equipped with an Intel® Core™ i7-11800H processor (8 physical cores), with the aerodynamic and control, linear hydrodynamic, second-order difference-frequency wave-load (Diff-QTF), and mooring calculations accounting for approximately 38%, 2%, 40%, and 19% of the total computational cost, respectively. Despite the additional cost associated with the second-order difference-frequency calculation, the overall simulation remained approximately twice as fast as real time, demonstrating the computational efficiency of the model for long-duration coupled simulations. For the subsequent comparison of response statistics and power spectral densities (PSDs), only the 420–3620 s interval was retained, excluding the initial and final portions of each time series. The experimental and numerical data were then resampled to a common time interval to ensure a consistent basis for comparison. The PSDs were calculated using MATLAB's pwelch function with a Hanning window and 50% overlap to obtain smoothed spectral estimates.

### *4.1. Wave–only cases*

Fig. 4 compares the standard deviations of platform surge, heave, pitch, and fore mooring-line tension obtained experimentally and from the three numerical models. The largest model dependence is observed for surge and mooring tension, where the linear model substantially underpredicts the measured response variability, particularly for steeper Waves 2, 3, and 5. Including second-order difference-frequency loads markedly improves the prediction, with the Diff-QTF model generally providing the closest agreement. For these three cases, the surge standard deviation (std) underprediction is reduced from 36.8–69.7% with the linear model to 10.4–18.2% with the Diff-QTF model, while the corresponding mooring-tension underprediction decreases from 29.7–50.9% to 6.1–13.9%. These cases are associated with relatively large wave heights and/or steepnesses, whereas the long-period, low-steepness Wave 4 shows much smaller model differences. For the mild Wave 1, the absolute responses remain small, although the surge and mooring predictions are still improved by including difference-frequency loads. In contrast, all three models provide similar heave predictions that generally agree well with the experimental results, indicating a relatively weak influence of difference-frequency loading and a response dominated by first-order wave-frequency excitation. For pitch, the linear model generally underpredicts the response by approximately 5–15% across Waves 1–5, while the inclusion of difference-frequency loads improves the agreement in most cases. The relative performance of Newman's approximation and the Diff-QTF model is case-dependent, with the latter slightly overpredicting the pitch response in Wave 2. Overall, accounting for second-order difference-frequency loads leads to a clear improvement in the prediction of surge motion and the associated mooring response, although some underprediction remains, particularly for Waves 3 and 5.

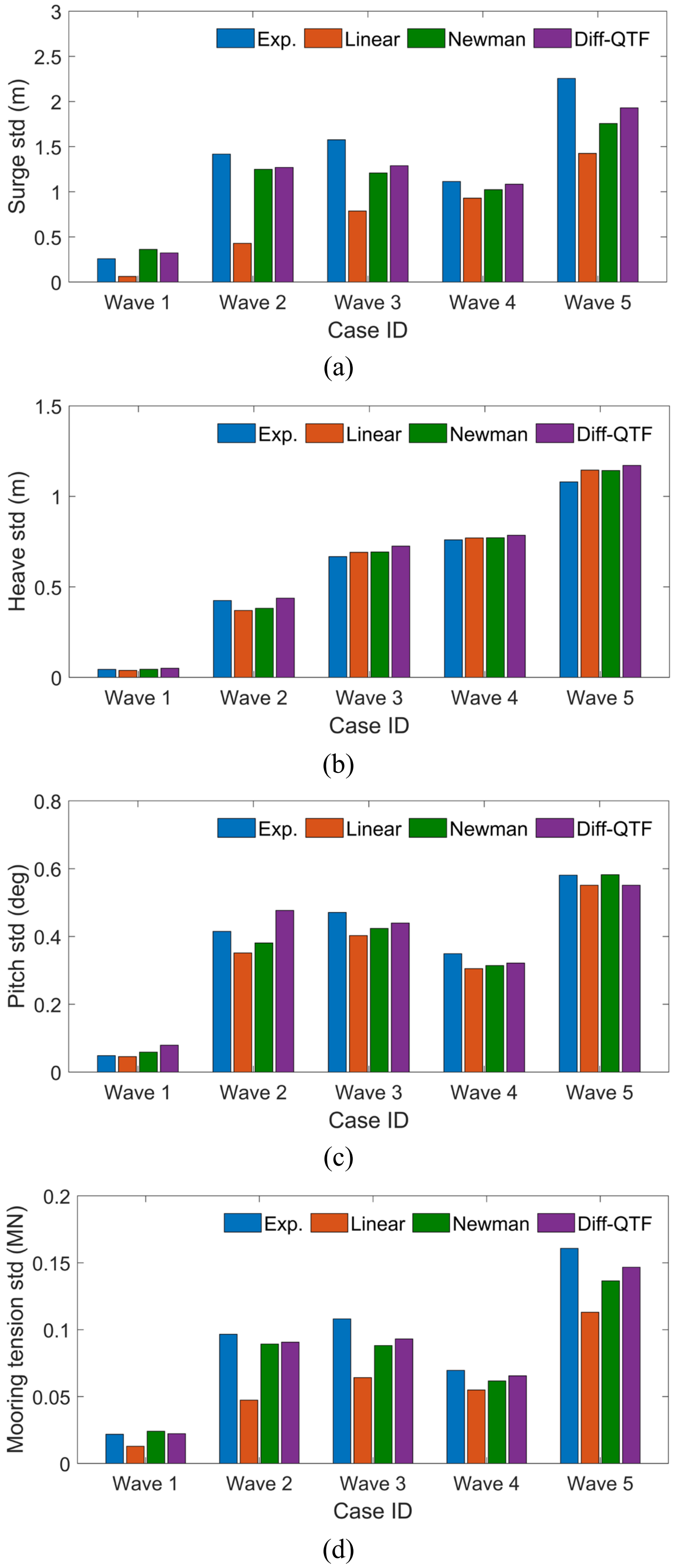

**Fig. 4.** Comparison of platform motion and mooring tension standard deviations between the experimental data and numerical results. (a) Surge; (b) Heave; (c) Pitch; (d) Fore mooring-line tension.

To further examine the frequency-dependent response, Waves 2 and 5 are selected for detailed spectral analysis as representative wave conditions. Fig. 5 presents the platform motion and mooring tension PSDs for Wave 2. Differences among the numerical models are mainly observed in the low-frequency range, while the three models give similar predictions and generally agree well with the experimental results over the main wave-frequency range of approximately 0.07–0.15 Hz. For surge, the experiment shows a pronounced low-frequency peak near the surge natural frequency (around 0.008 Hz), which is substantially underpredicted by the linear model. The inclusion of difference-frequency wave loads markedly increases the low-frequency spectral energy, particularly around the surge natural frequency, with the Newman and Diff-QTF models showing much closer agreement with the experiment. The fore mooring-line tension exhibits a similar behaviour, consistent with its close coupling with the low-frequency surge motion. The spectral results indicate that the improvements in surge and mooring-tension standard deviations achieved by the Newman and Diff-QTF models, as shown in Fig. 4, are primarily attributable to a more accurate representation of the low-frequency response. For heave, the main model differences occur near the heave natural frequency (approximately 0.05 Hz), where the inclusion of difference-frequency wave loads improves the prediction, with the Diff-QTF model providing the closest agreement with the experiment. For pitch, the model differences are concentrated near the natural-frequency peak (approximately 0.033 Hz), where the Diff-QTF model predicts a stronger response and slightly overpredicts the experimental peak magnitude. Overall, the largest model differences tend to occur near the natural frequencies of the respective platform motions, where differences in the predicted low-frequency excitation can be amplified by resonance.

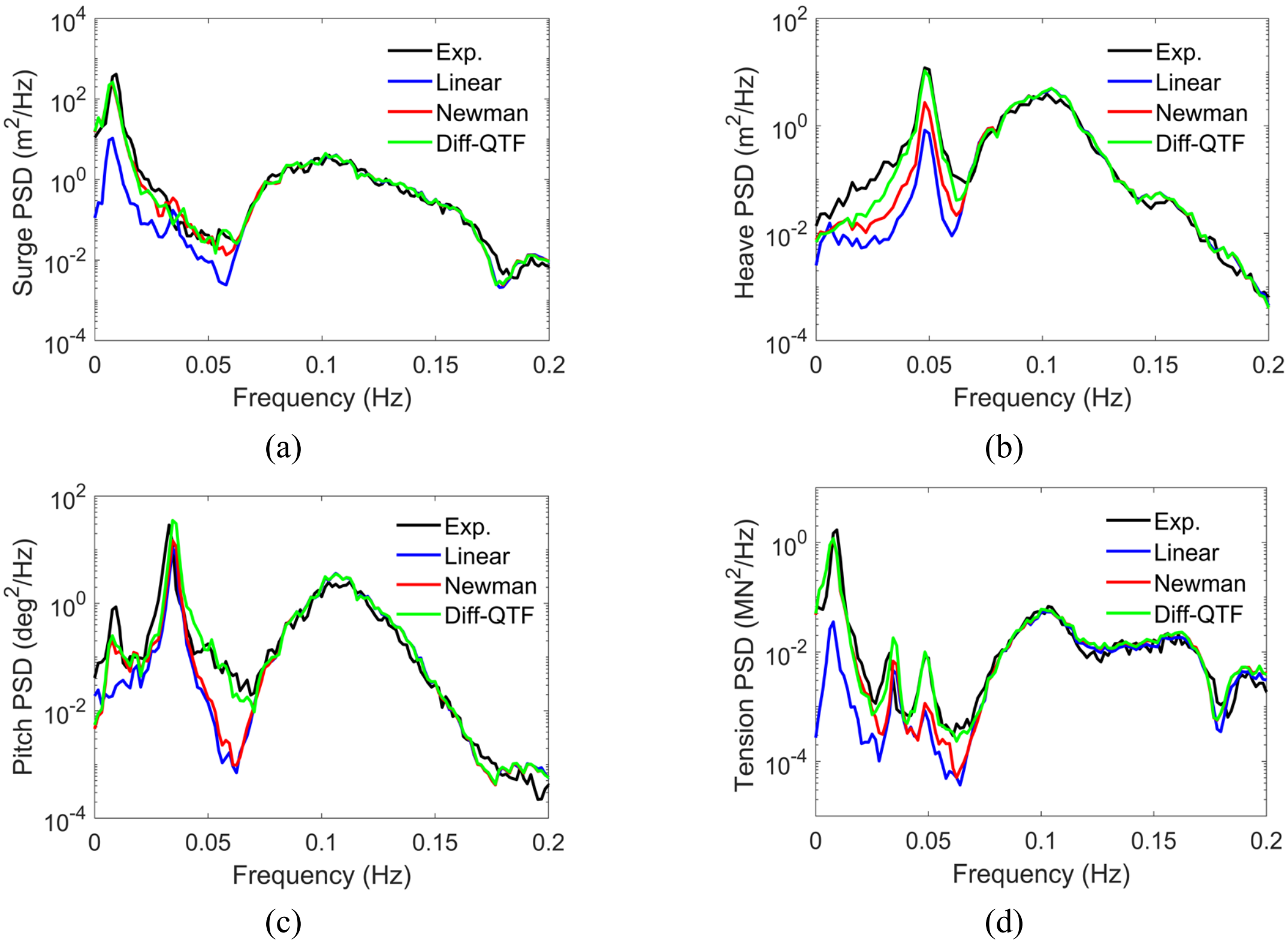

**Fig. 5.** Comparison of platform motion and mooring tension PSDs between the experimental data and numerical results for Wave 2. (a) Surge PSD; (b) Heave PSD; (c) Pitch PSD; (d) Fore mooring-line tension PSD.

Fig. 6 presents the corresponding PSDs for Wave 5, representing the severe sea state under the parked condition and characterised by relatively large-amplitude, long-period waves. Compared with Wave 2, more noticeable discrepancies between the numerical predictions and the experiment remain near the platform natural frequencies despite the inclusion of difference-frequency wave loads. For surge, the Newman and Diff-QTF models substantially improve the low-frequency response but still underpredict the experimental peak, with a similar tendency observed for the fore mooring-line tension. Heave is generally well predicted, although some discrepancy remains near the heave natural frequency, while the pitch natural-frequency peak is also underpredicted. Notably, local wave breaking, green-water events, and slamming-like impacts were observed around the platform during the Wave 5 tests, indicating the presence of strongly nonlinear hydrodynamic effects. These observations, together with the remaining underprediction of the low-frequency response in Wave 5 and the similar underprediction observed in Wave 3, suggest that potential-flow difference-frequency excitation alone may not fully account for the measured response under these conditions. Similar underprediction of low-frequency response has been reported for semi-submersible FOWTs, with resonant behaviour shown to be sensitive to both motion-dependent viscous damping and the representation of nonlinear hydrodynamic excitation at low frequencies (Li and Bachynski-Polić, 2021; Wang et al., 2022). In the present model, motion-dependent viscous damping is represented through empirical quadratic drag terms, whereas wave-induced viscous nonlinearities associated with fluid–structure relative motion and strongly nonlinear free-surface interactions are not explicitly represented. These effects may become more pronounced under large-amplitude wave conditions, providing additional low-frequency excitation beyond that captured by the potential-flow QTFs and potentially contributing to the remaining discrepancies in Waves 3 and 5.

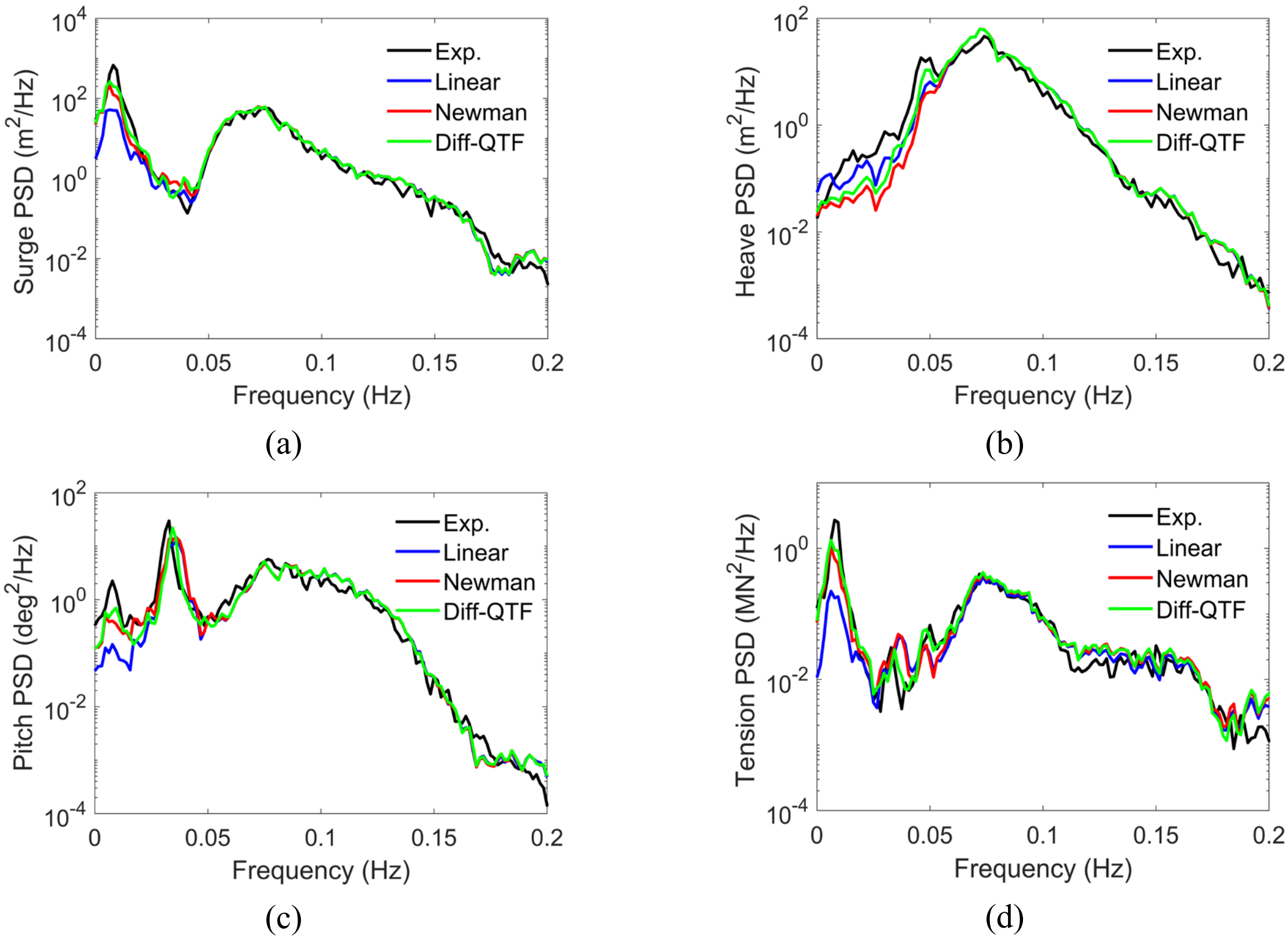

**Fig. 6.** Comparison of platform motion and mooring tension PSDs between the experimental data and numerical results for Wave 5. (a) Surge PSD; (b) Heave PSD; (c) Pitch PSD; (d) Fore mooring-line tension PSD.

Figs. 7 and 8 show the empirical exceedance probability distributions of peak amplitudes for platform motions and mooring tension in Waves 2 and 5, respectively, with the exceedance probability plotted on a logarithmic scale to better resolve the low-probability tail. The peak amplitude is defined as the absolute deviation of each local extremum from the corresponding mean value. For all datasets, local maxima and minima were identified using a minimum peak distance of 2.0 s and minimum peak prominences of 0.05 m for surge and heave, 0.03° for pitch, and 0.04 MN for mooring tension, to consistently suppress small-scale fluctuations and exclude insignificant local extrema. In Wave 2, the linear model substantially underpredicts the surge and fore mooring-tension peaks over almost the entire exceedance range, while the Newman and Diff-QTF models provide marked improvements. For heave, the Diff-QTF model gives the closest agreement with the experiment, particularly for the larger peaks. For pitch, the Newman model improves upon the linear prediction, while the Diff-QTF model predicts somewhat larger peak amplitudes, consistent with the slight overprediction near the pitch natural frequency shown in Fig. 5(c). In Wave 5, the surge peak amplitudes predicted by the linear model remain well below the experimental distribution. Both the Newman and Diff-QTF models improve the prediction over the intermediate exceedance range, although some underprediction remains, with the Diff-QTF model providing closer agreement. Towards the extreme tail, however, the Diff-QTF model predicts slightly larger peak amplitudes than observed experimentally. A broadly similar pattern is observed for the fore mooring-line tension, where the Diff-QTF model improves the prediction over most of the distribution but tends to overpredict the largest peaks towards the extreme tail. For heave, all three models reproduce the experimental distribution reasonably well, with only small differences among the prediction. For pitch, the agreement is also generally good, although the largest peaks are underpredicted towards the extreme tail.

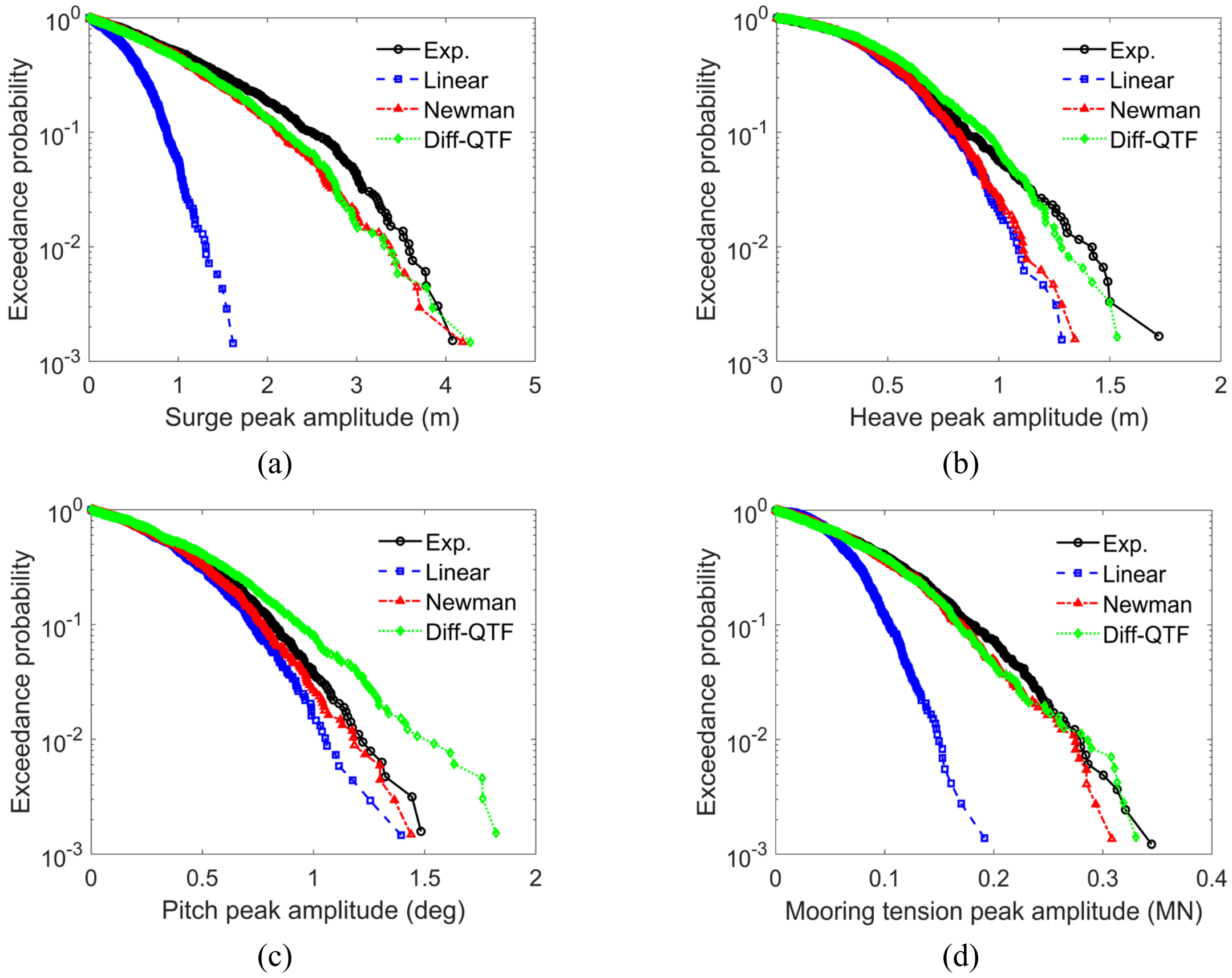

**Fig. 7.** Exceedance probability distributions of peak amplitudes for platform motions and mooring tension in Wave 2. (a) Surge; (b) Heave; (c) Pitch; (d) Fore mooring-line tension.

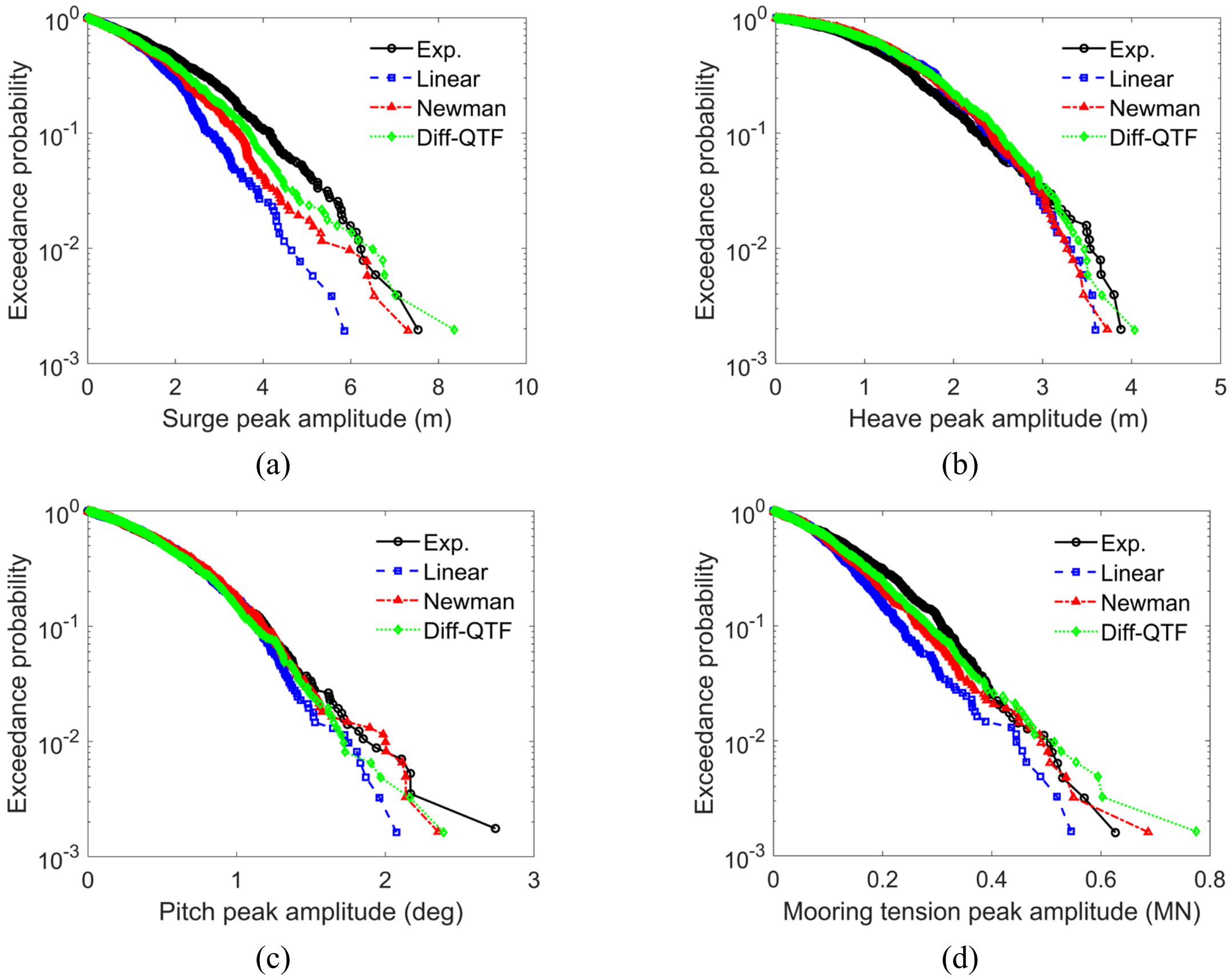

**Fig. 8.** Exceedance probability distributions of peak amplitudes for platform motions and mooring tension in Wave 5. (a) Surge; (b) Heave; (c) Pitch; (d) Fore mooring-line tension.

### *4.2. Combined wind–wave cases*

Fig. 9 compares the standard deviations of platform surge, heave, pitch, and fore mooring-line tension obtained experimentally and from the three numerical models for the combined cases. Unlike the wave-only cases, where the inclusion of difference-frequency loads leads to clear improvements in surge and mooring-response predictions across several wave conditions, the differences among the three models are generally small for the combined U8–U14 cases, with all models showing reasonable agreement with the experiments. For these cases, the maximum relative errors in the surge and mooring-tension standard deviations remain within 13.7% and 14.8% for the linear model, and within 7.3% and 4.4% for the Diff-QTF model, respectively. More noticeable differences emerge only in the U20 cases, where the linear model gives the lowest responses and the Newman and Diff-QTF models increase the predicted variability towards the experimental values. For these cases, the maximum underprediction is reduced from 46.6% to 17.8% for surge and from 38.6% to 13.4% for mooring tension with the Diff-QTF model. Heave is predicted similarly by the three models across most cases, although the Diff-QTF model tends to give slightly larger responses. For pitch, the differences among the three models remain small, while the numerical predictions generally fall below the experimental values for U8–U14 and become closer at U20.

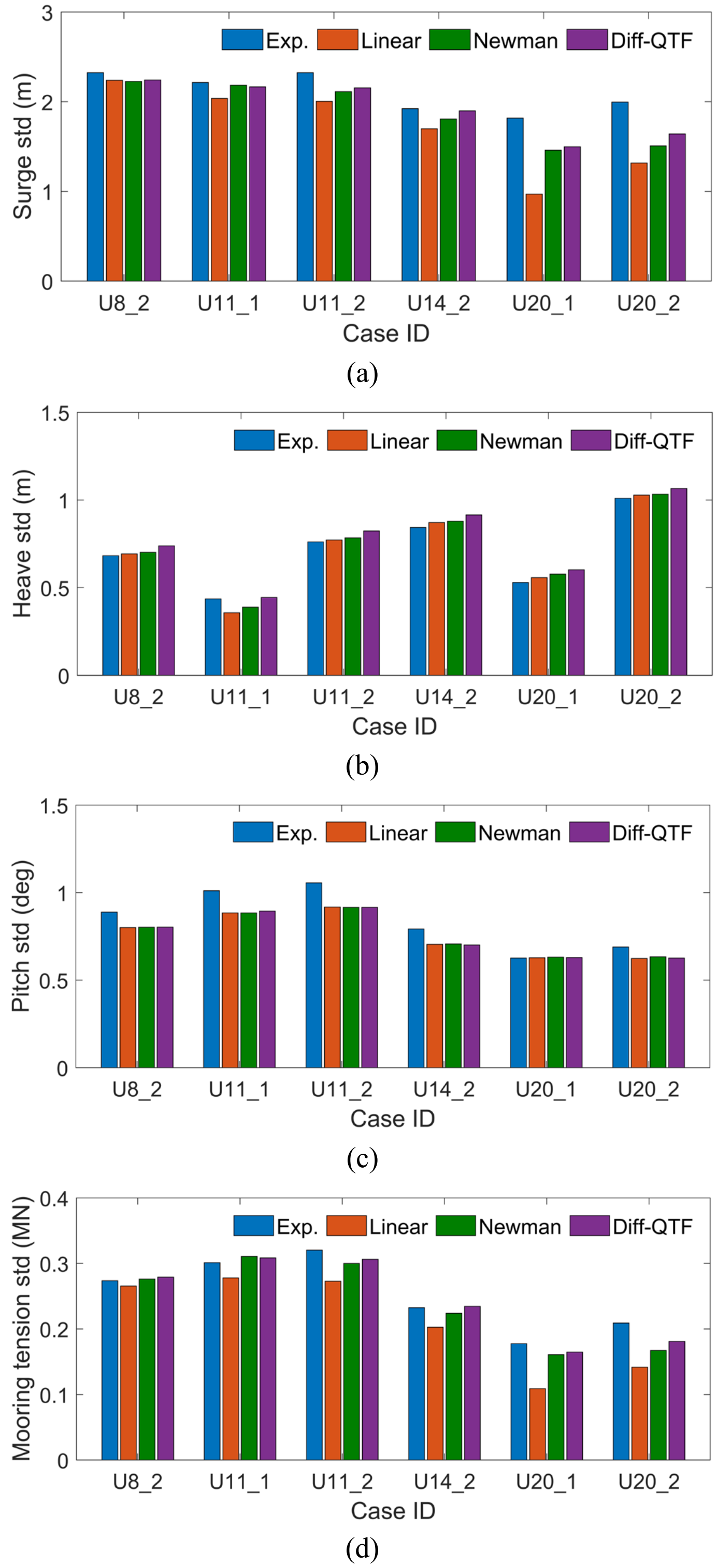

**Fig. 9.** Comparison of platform motion and mooring tension standard deviations between the experimental data and numerical results for the combined DLC 1.6 cases. (a) Surge; (b) Heave; (c) Pitch; (d) Fore mooring-line tension.

The reduced sensitivity to the hydrodynamic model in the combined cases is likely associated with the additional low-frequency aerodynamic excitation from turbulent winds, which reduces the relative contribution of difference-frequency wave loading to the overall response. To assess the relative magnitudes of the low-frequency aerodynamic and hydrodynamic forcing, the surge-direction thrust fluctuations and thrust-induced pitch moments are compared with the corresponding difference-frequency surge forces and pitch moments using the Diff-QTF model. The surge-direction thrust component is evaluated in the global surge direction, and the mean values are removed from all forcing components before the spectral analysis. Figs. 10 and 11 compare the PSDs of surge force and pitch moment respectively, for the representative U11_2 and U20_2 cases. Around the surge natural frequency, the thrust-induced component is markedly larger than the difference-frequency surge force at U11_2, whereas at U20_2 the latter becomes comparable to or exceeds the thrust-induced component. For pitch, the contrast between the thrust-induced and difference-frequency contributions is generally more pronounced than for surge, owing to the large moment arm of the rotor thrust. Nevertheless, the difference-frequency contribution becomes relatively more important at U20_2.

These spectral differences highlight the case-dependent relative importance of thrust-induced and difference-frequency forcing. To quantify their relative magnitudes over the relevant low-frequency ranges, band-limited root-mean-square (RMS) values are calculated over 0.002–0.025 Hz for surge force and 0.002–0.05 Hz for pitch moment. These ranges were selected to cover the low-frequency regions most relevant to the surge and pitch responses, including the substantial pitch response below its natural frequency associated with turbulent wind excitation (Zhang et al., 2026b), while excluding the main wave-frequency region. These frequency bands are indicated by the light-blue shaded regions in Figs. 10 and 11, while the dashed lines denote the corresponding platform natural frequencies. The band-limited RMS is defined as:

$$X_{\mathrm{RMS}} = \sqrt{\int_{f_1}^{f_2} S_X(f)\,\mathrm{d}f} \tag{21}$$

where $S_X(f)$ is the PSD of the corresponding force or moment, and $f_1$ and $f_2$ are the lower and upper bounds of the selected frequency band. The resulting RMS values are evaluated for all cases considered and compared in Fig. 12 to quantify the relative magnitudes of the thrust-induced and difference-frequency contributions to surge force and pitch moment. For surge, thrust-induced forcing dominates from U8 to U14, with RMS ratios of roughly 2–3.5, while the thrust-induced RMS generally decreases with increasing wind speed and the difference-frequency contribution becomes relatively more important, eventually exceeding the thrust-induced component at U20. For pitch, the thrust-induced moment remains dominant across all cases, although its RMS ratio to the difference-frequency contribution decreases from about 7–8 at U8–U11 to around 2 at U20. These trends are consistent with the forcing PSD comparisons and indicate an increasing relative importance of difference-frequency hydrodynamic forcing towards U20, where stronger wave conditions are accompanied by weaker thrust fluctuations. The corresponding platform-motion and mooring-tension responses are examined further through their PSDs for representative combined cases.

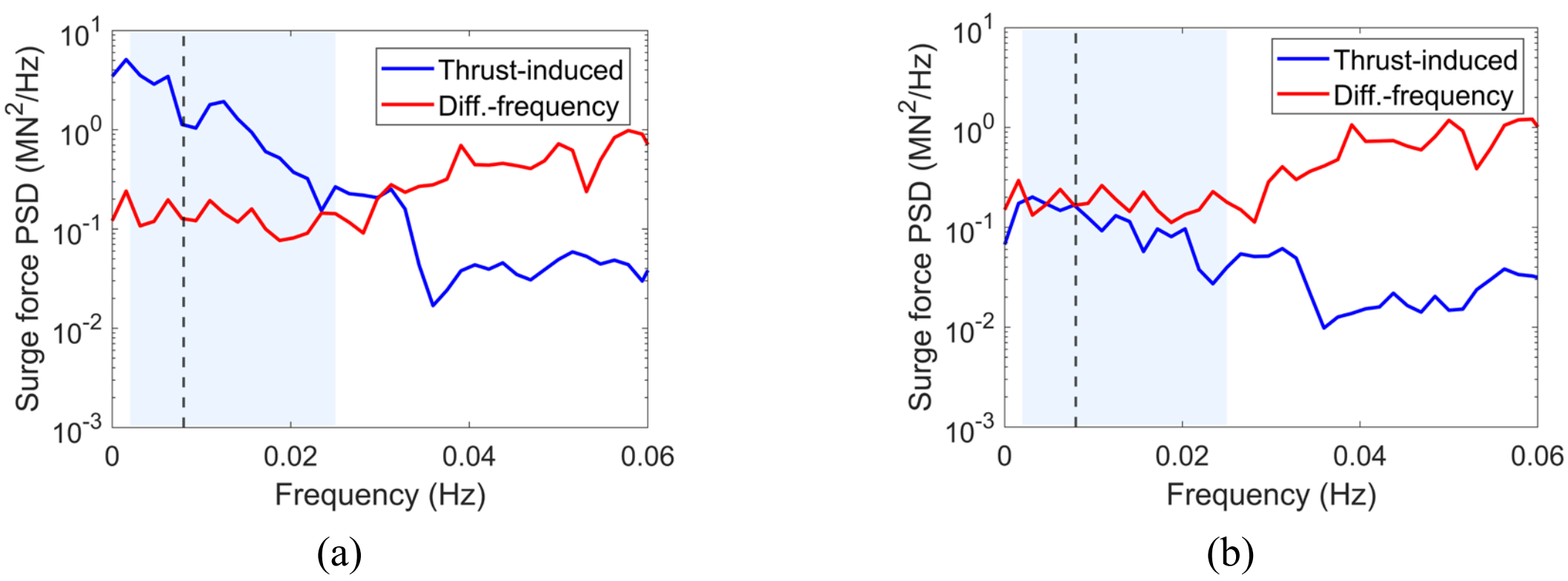

**Fig. 10.** Comparison of low-frequency PSDs of surge-direction thrust fluctuations and difference-frequency surge force for different DLC 1.6 cases. (a) U11_2; (b) U20_2. The dashed lines indicate the surge natural frequency (0.008 Hz), while the light-blue shaded regions denote the frequency range (0.002–0.025 Hz) used to calculate the band-limited RMS values.

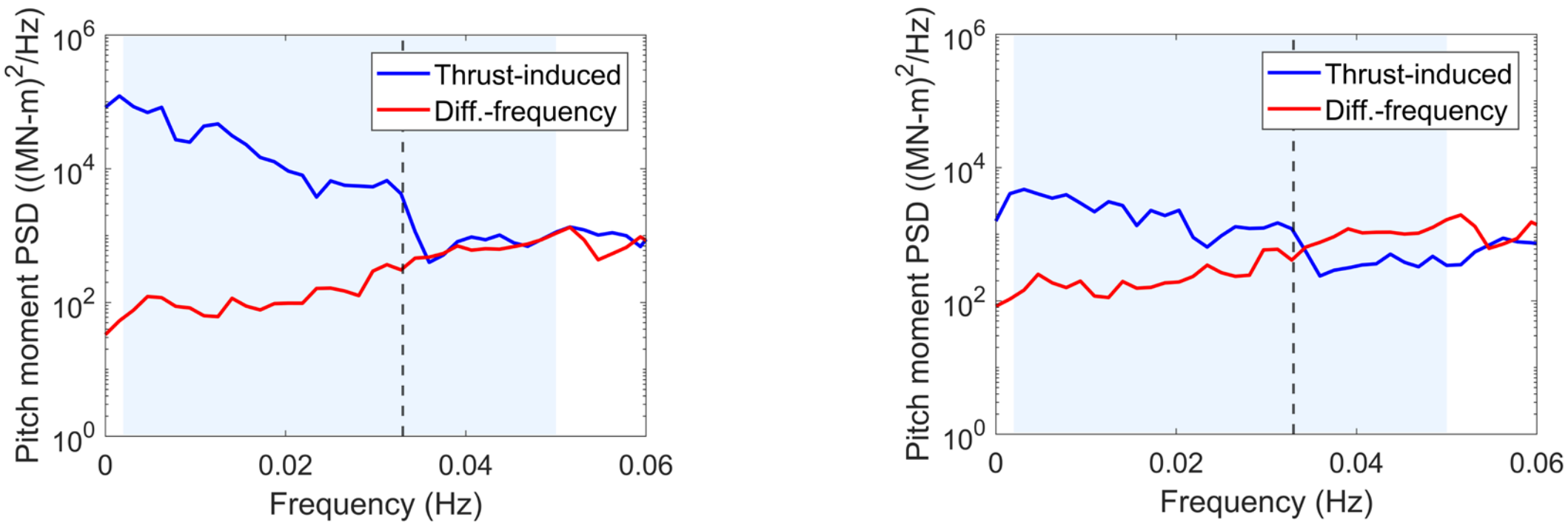

(a) (b)

**Fig. 11.** Comparison of low-frequency PSDs of thrust-induced and difference-frequency pitch moment for different DLC 1.6 cases. (a) U11_2; (b) U20_2. The dashed lines indicate the pitch natural frequency (0.033 Hz), while the light-blue shaded regions denote the frequency range (0.002–0.05 Hz) used to calculate the band-limited RMS values.

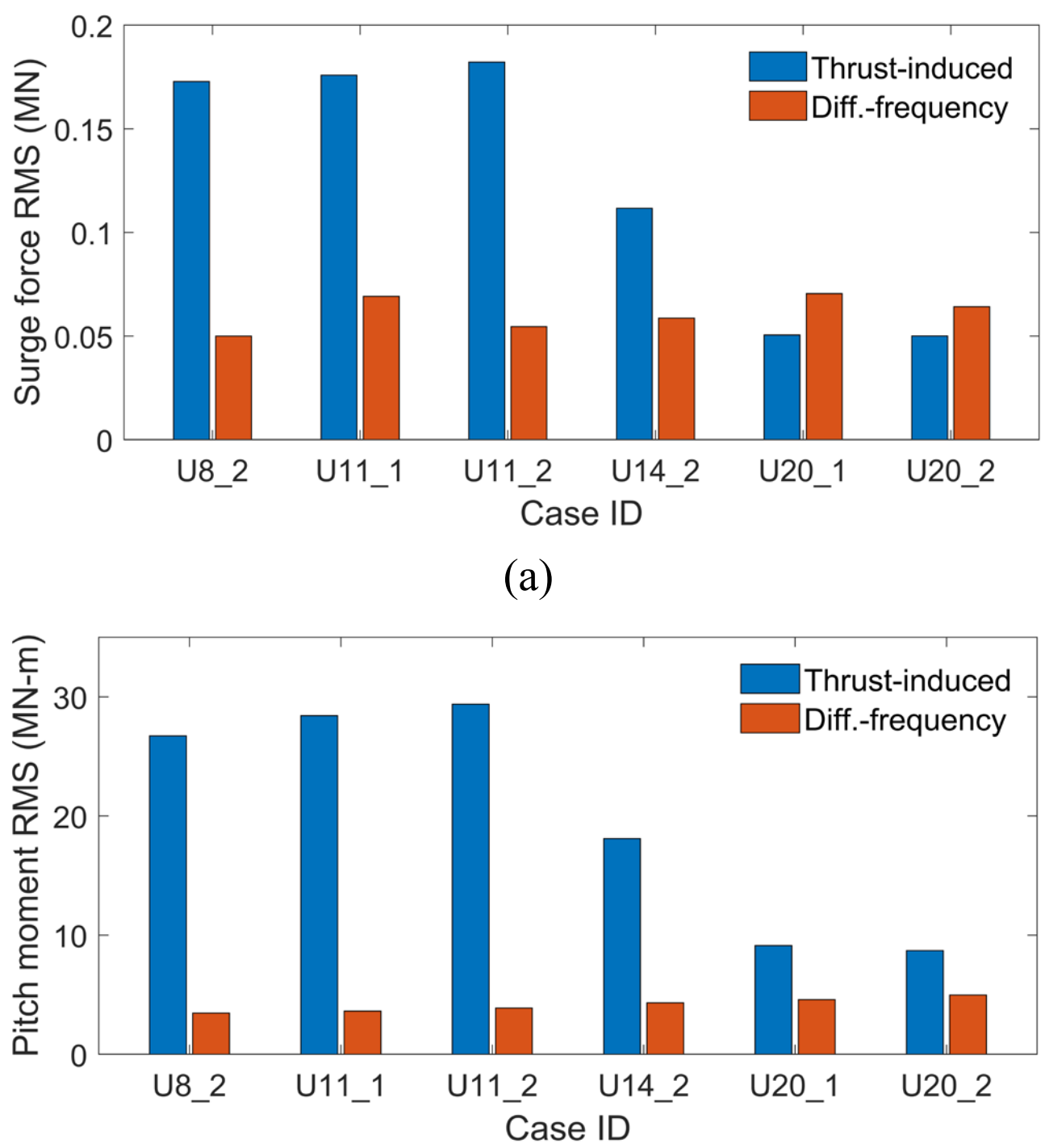

(a)

(b)

**Fig. 12.** Comparison of band-limited RMS values of thrust-induced and difference-frequency forcing for the combined DLC 1.6 cases. (a) Surge force; (b) Pitch moment.

Fig. 13 compares the surge PSDs for four representative DLC 1.6 cases with different wind speeds. Across all cases, the three numerical models give very similar predictions over the main wave-frequency range and generally agree well with the experimental spectra. The most pronounced model differences occur in the low-frequency region. For U8_2 and U11_2, the low-frequency responses are broadly comparable, although some separation among the models is visible around 0.03–0.05 Hz, where the spectral amplitudes remain small relative to the dominant wind- and wave-induced response regions. The differences become more evident at U14_2 and especially U20_2, where the linear model underpredicts the low-frequency peak, while the Newman and Diff-QTF models predict stronger low-frequency responses, with the Diff-QTF model providing the closest agreement with the experiment. This trend is consistent with the increasing relative importance of difference-frequency hydrodynamic forcing towards the higher wind-speed cases.

However, the remaining discrepancy in the surge PSD at U20_2 suggests that additional nonlinear hydrodynamic effects beyond potential-flow difference-frequency loading may also contribute under the stronger wave condition. The three hydrodynamic models produce very similar pitch PSDs, consistent with the dominant low-frequency contribution of the thrust-induced moment. Compared with the wave-only cases, the combined cases exhibit substantially greater low-frequency pitch energy below about 0.04 Hz due to turbulent wind excitation, as also observed experimentally in Zhang et al. (2026b). For heave, the Diff-QTF model gives the closest agreement near the heave natural frequency (0.05 Hz), although this has little influence on the overall response variability. The corresponding pitch and heave PSDs are therefore omitted for brevity. The fore mooring-line tension PSDs in Fig. 14 show a similar low-frequency trend to surge, with differences among the hydrodynamic models becoming more evident towards U20. The Newman and Diff-QTF models increase the low-frequency tension response relative to the linear model, with the Diff-QTF model providing the closest agreement with the experiment. The Diff-QTF model also improves the tension prediction around 0.05 Hz, consistent with the improved heave prediction near its natural frequency. Outside the low-frequency range, the three models give broadly similar spectra and agree well with the experimental measurements.

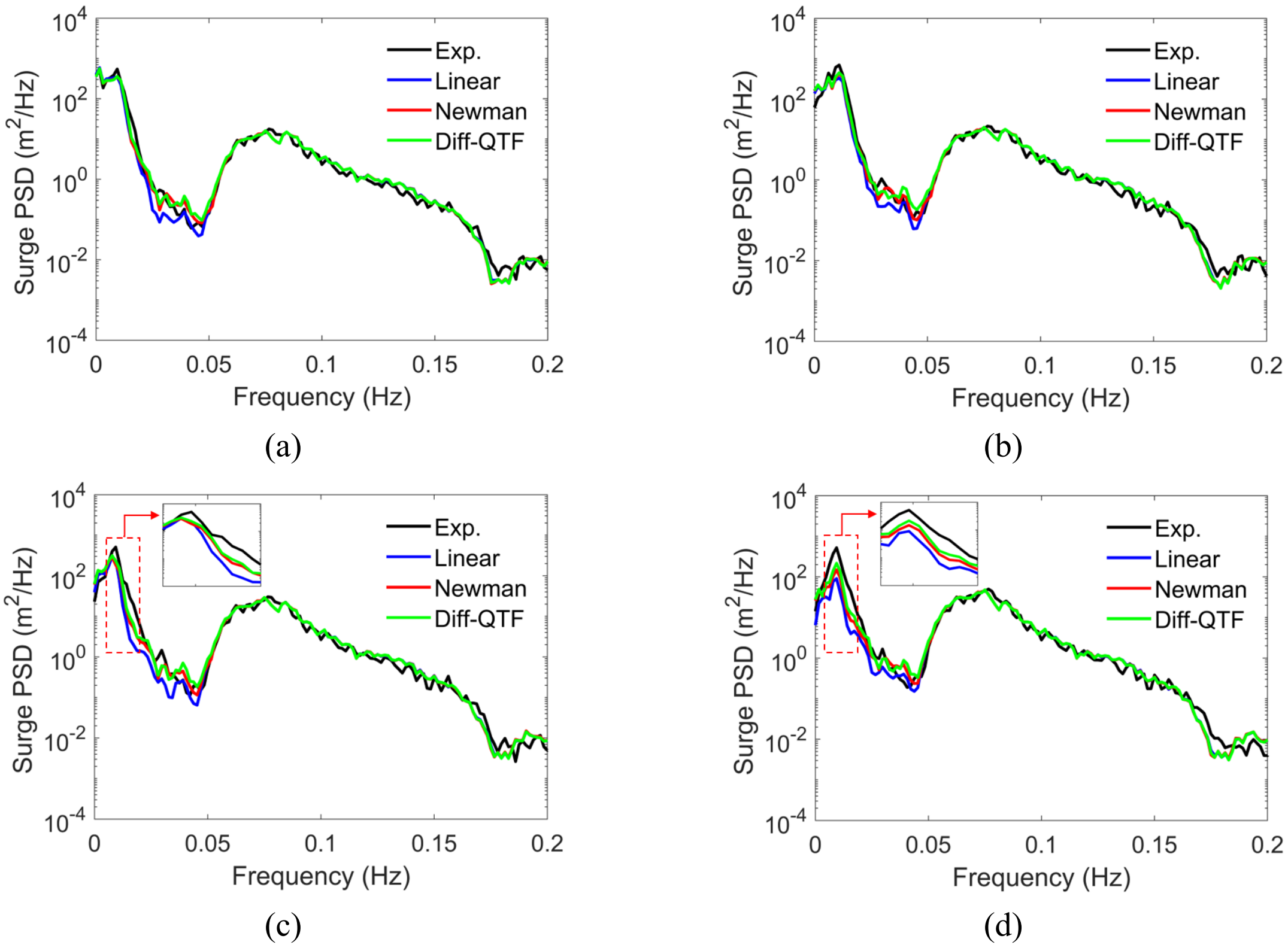

**Fig. 13.** Platform surge PSDs for different DLC 1.6 cases. (a) U8_2; (b) U11_2; (c) U14_2; (d) U20_2.

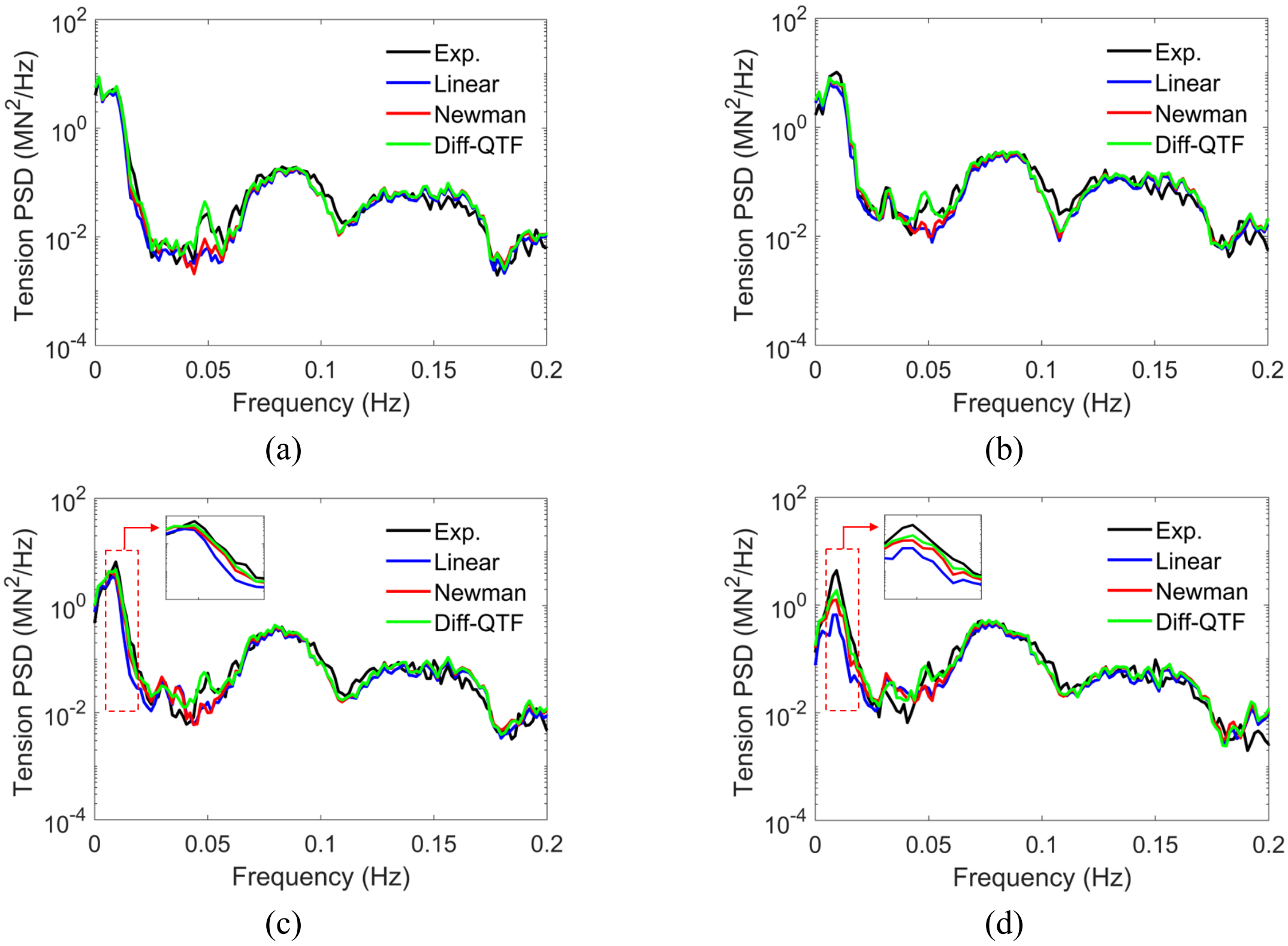

**Fig. 14.** Fore mooring-line tension PSDs for different DLC 1.6 cases. (a) U8_2; (b) U11_2; (c) U14_2; (d) U20_2.

Fig. 15 further compares the empirical exceedance probability distributions of surge peak amplitudes for the same representative cases. The differences among the numerical models are relatively small at U8_2 but become more evident from U11_2 onwards, particularly in the larger-amplitude range. Including difference-frequency loading generally shifts the predicted distributions towards the measurements, with the Diff-QTF model providing the closest agreement over much of the exceedance range, although some underprediction remains for the higher-amplitude peaks due to additional nonlinear hydrodynamic effects, especially at U20_2. A similar case dependence is observed for the fore mooring-line tension peak amplitudes in Fig. 16, consistent with their close coupling with surge motion. At U8_2, the three numerical models remain close over most of the exceedance range and show good agreement with the measurements. At U11_2 and U14_2, including difference-frequency loading improves the prediction of the larger peaks, with the Diff-QTF model generally giving the closest agreement. At U20_2, the Newman and Diff-QTF models improve agreement over the intermediate exceedance range, while the Diff-QTF model tends to overpredict the extreme tail.

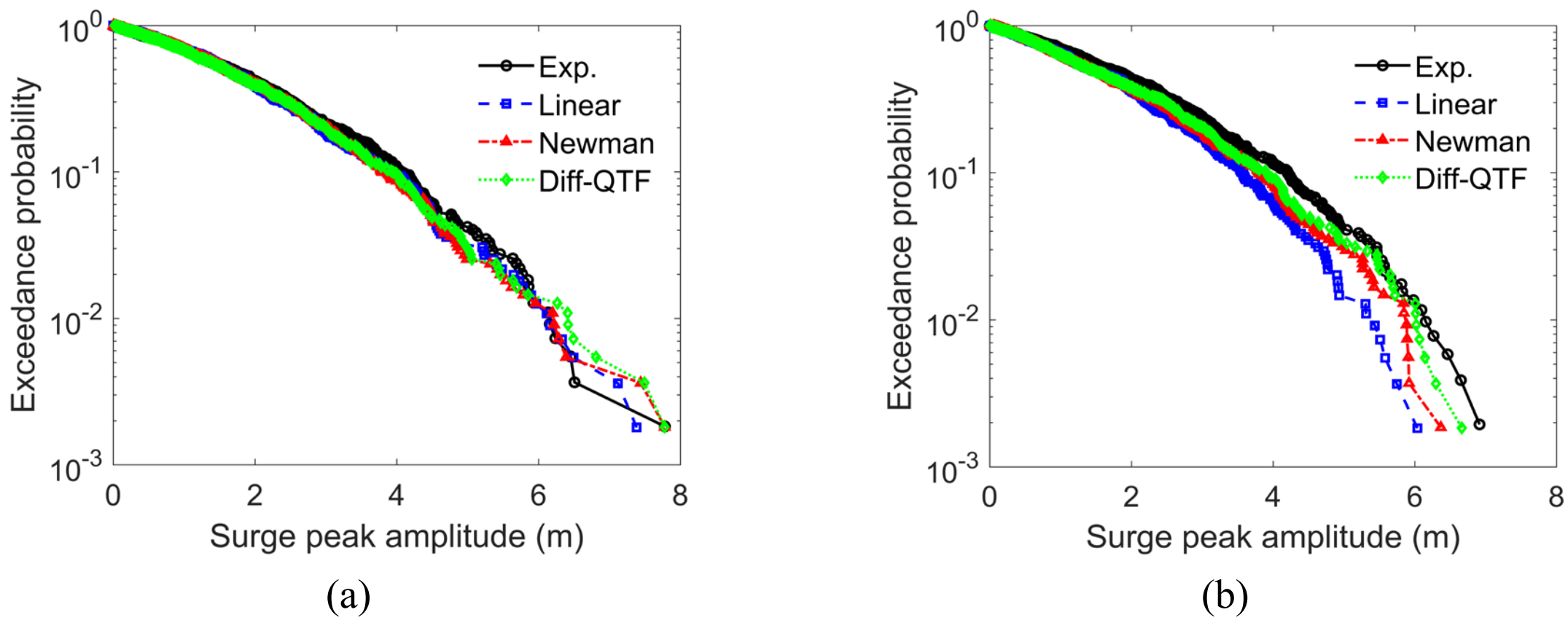

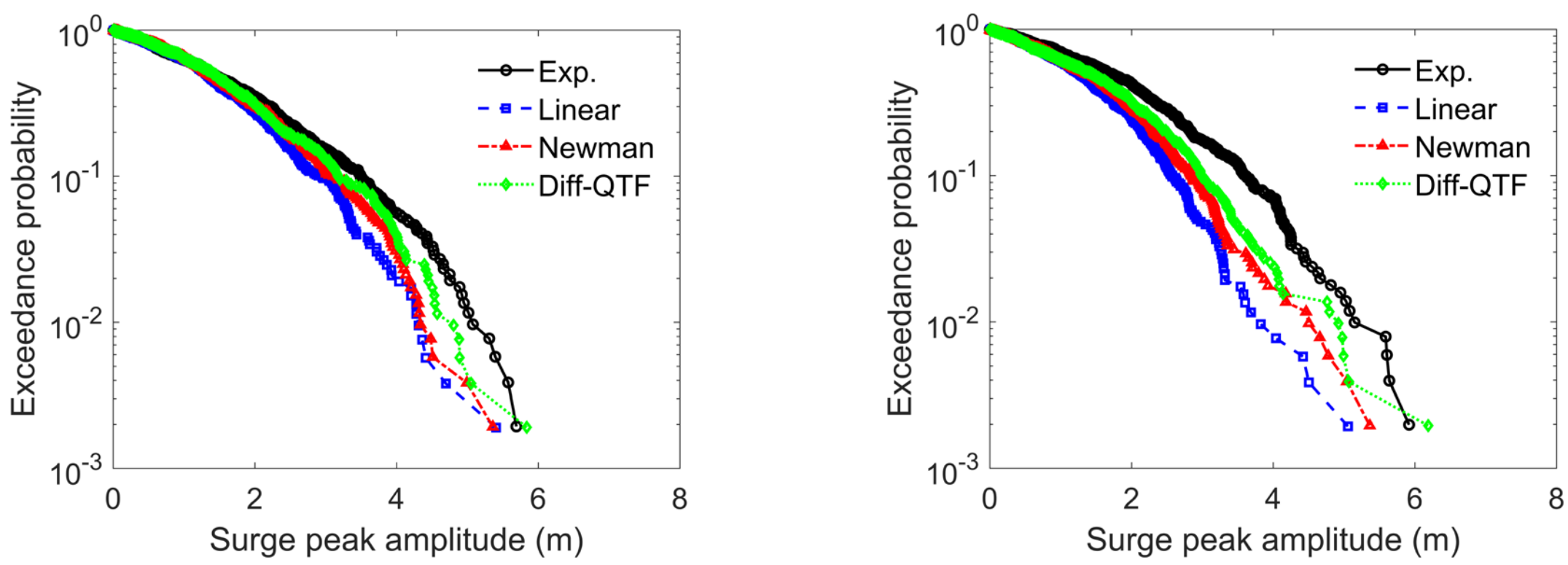

(c) (d)

**Fig. 15.** Exceedance probability distributions of platform surge peak amplitudes for different DLC 1.6 cases. (a) U8_2; (b) U11_2; (c) U14_2; (d) U20_2.

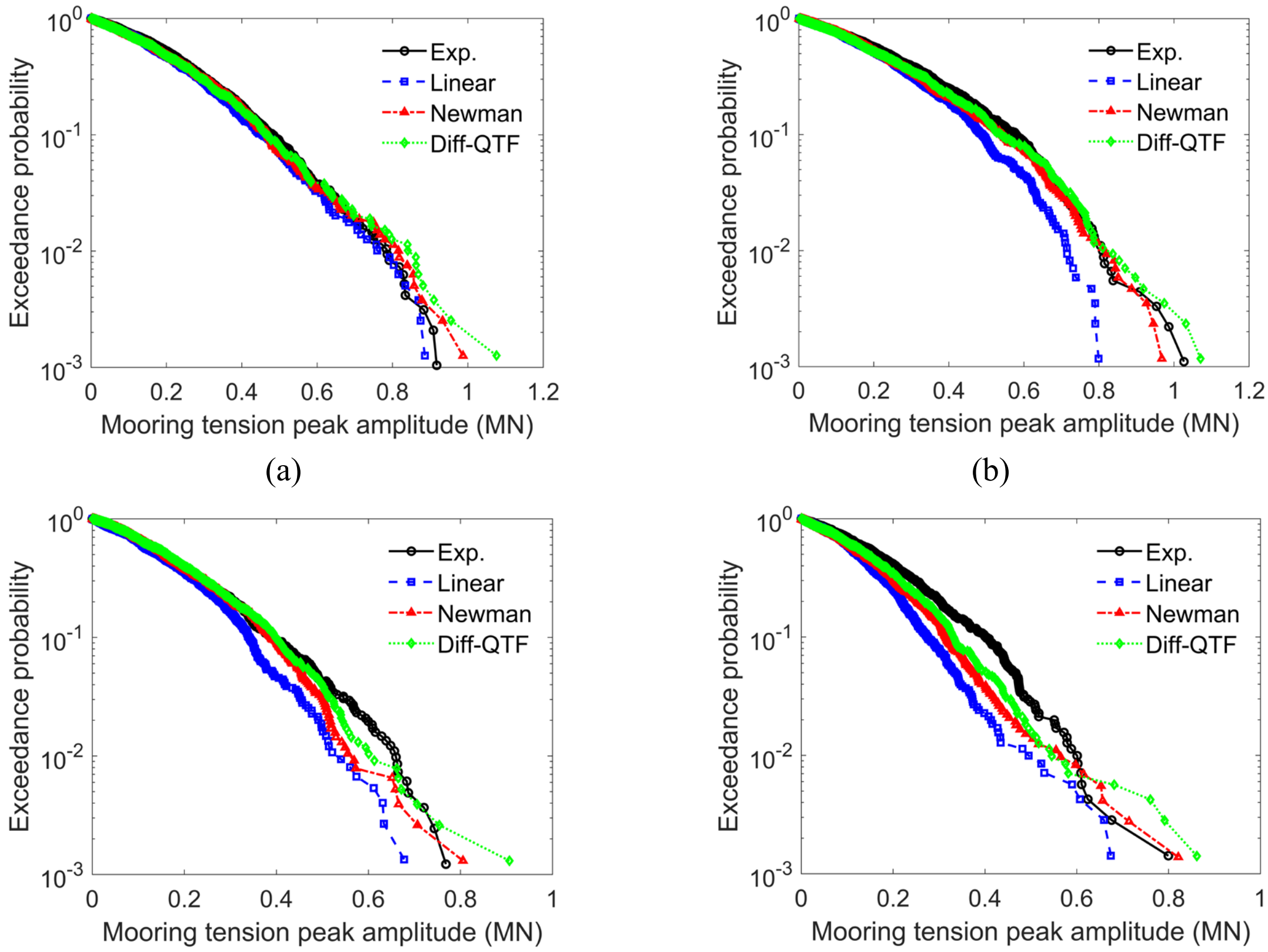

(c) (d)

**Fig. 16.** Exceedance probability distributions of fore mooring-line tension peak amplitudes for different DLC 1.6 cases. (a) U8_2; (b) U11_2; (c) U14_2; (d) U20_2.

### 4.3. *Effect of simulation duration*

Since the available wave-basin measurements cover approximately 1 h at full scale, the numerical–experimental comparisons in the preceding sections were based on matched 1-h records. To assess the influence of simulation duration on the response characteristics predicted by the Diff-QTF model, two representative combined wind–wave cases were selected: DLC1.6_U11_2 and DLC1.6_U20_2. For each case, a single continuous numerical simulation was used, from which progressively longer response records of 1 h, 3 h, and 5 h were extracted after removing the initial transient. The three records share the same starting point, allowing the effect of increasing record length to be assessed within the same stochastic realisation without introducing additional variability from different random realisations. The 5-h results are used as the reference for evaluating the sensitivity to record length.

The influence of simulation duration was first examined through the standard deviations of the platform motions and fore mooring-line tension, as summarised in Tables 5 and 6. For U11_2, the 1-h results differ from the 5-h values by at most 1.93%, while the corresponding maximum difference for the 3-h results is 1.23%. For U20_2, the maximum differences are 2.03% and 0.64% for the 1-h and 3-h records, respectively. These small variations are likely attributable to sampling variability and indicate limited sensitivity of the overall response statistics to record length. The corresponding PSDs (not shown) also exhibit only minor differences among the three record lengths.

**Table 5.** Effects of simulation duration on response statistics with the Diff-QTF model for DLC1.6_U11_2. The values in parentheses are relative errors with respect to 5 h simulated results.

| Simulation duration | Surge std (m) | Heave std (m) | Pitch std (deg) | Mooring tension std (MN) |
|---|---|---|---|---|
| 1 h | 2.0167 (−1.93%) | 0.8251 (+0.41%) | 0.8851 (−0.90%) | 0.2964 (−1.21%) |
| 3 h | 2.0637 (+0.36%) | 0.8192 (−0.31%) | 0.9040 (+1.23%) | 0.3012 (+0.38%) |
| 5 h | 2.0563 | 0.8218 | 0.8931 | 0.3001 |

**Table 6.** Effects of simulation duration on response statistics with the Diff-QTF model for DLC1.6_U20_2. The values in parentheses are relative errors with respect to 5 h simulated results.

| Simulation duration | Surge std (m) | Heave std (m) | Pitch std (deg) | Mooring tension std (MN) |
|---|---|---|---|---|
| 1 h | 1.6891 (+0.76%) | 1.0694 (+0.13%) | 0.6555 (−0.26%) | 0.1893 (+2.03%) |
| 3 h | 1.6709 (−0.32%) | 1.0612 (−0.64%) | 0.6540 (−0.49%) | 0.1856 (+0.03%) |
| 5 h | 1.6763 | 1.0680 | 0.6572 | 0.1855 |

The effect of record length on the larger response excursions was further examined using the empirical exceedance distributions of peak amplitudes. Fig. 17 compares the 1-h, 3-h, and 5-h results for U11_2, with 95% confidence intervals estimated using 2000 peak-wise bootstrap resamples. The exceedance probability is plotted on a logarithmic scale to better resolve the low-probability tail, where the largest response peaks occur. Although the logarithmic scaling visually expands the tail region, it contains only a small fraction of the ranked peaks. The three records therefore produce very similar distributions for the vast majority of the peaks, particularly for heave, while more noticeable differences emerge towards the low-probability tail. The bootstrap confidence intervals are generally narrow over the main part of the distributions but widen towards the low-probability tail, reflecting the greater sampling uncertainty associated with the relatively small number of extreme peaks. Nevertheless, the longer records capture additional large-amplitude events that are absent from the 1-h interval, most notably for surge and fore mooring-line tension. Pitch shows a similar but less pronounced sensitivity in the low-probability tail. Overall, these results highlight the benefit of longer records for characterising relatively rare, large-amplitude response events with reduced sampling uncertainty.

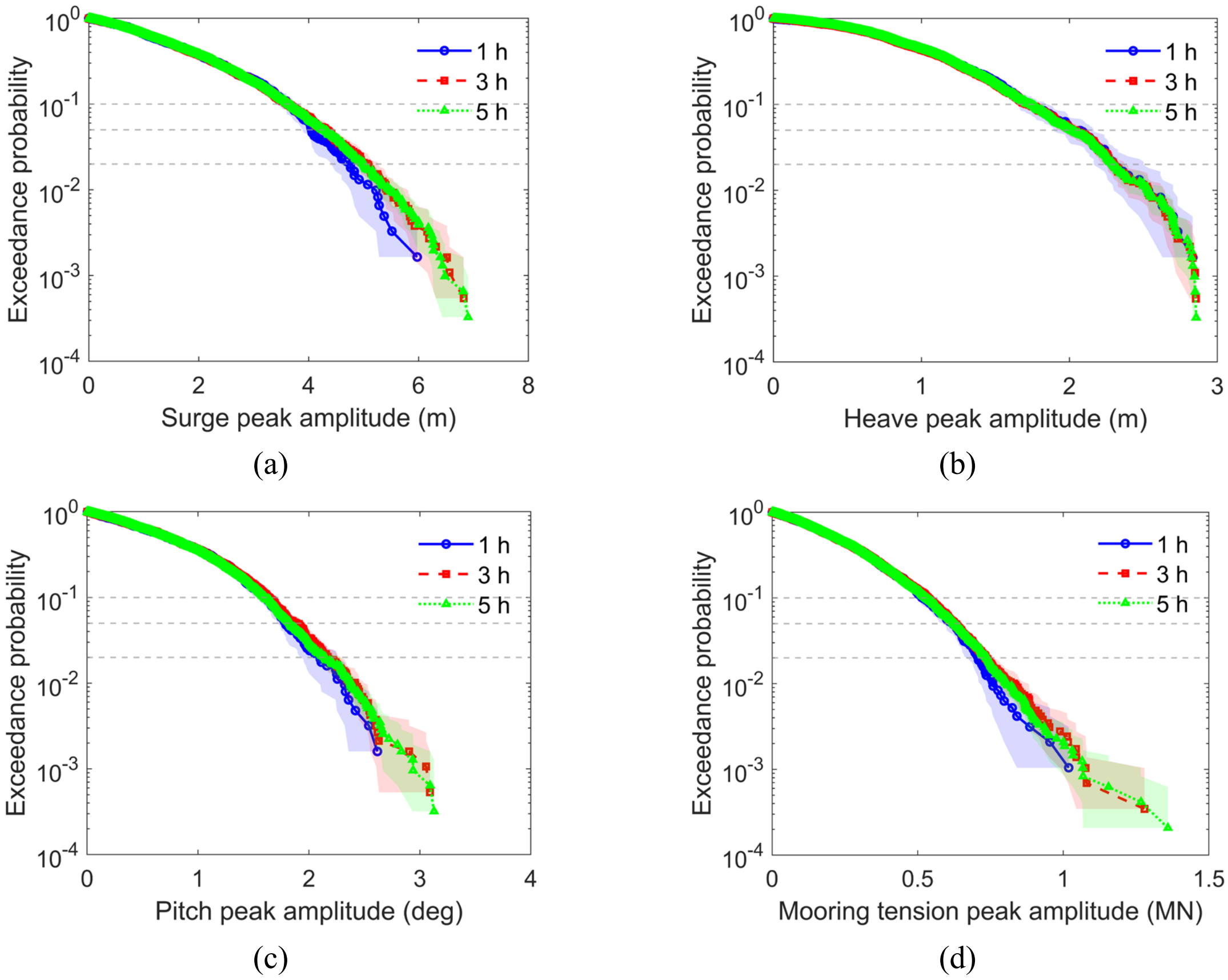

**Fig. 17.** Exceedance probability distributions of peak amplitudes for different simulation durations in DLC1.6_U11_2. (a) Surge; (b) Heave; (c) Pitch; (d) Fore mooring-line tension. The grey dashed lines denote the selected exceedance probabilities of $P_{\mathrm{ex}} = 0.10$, 0.05, and 0.02. The shaded regions represent the 95% bootstrap confidence intervals.

Figs. 18 and 19 provide quantitative comparisons of the peak amplitudes at the selected exceedance probabilities $P_{\mathrm{ex}} = 0.10$, 0.05, and 0.02 for U11_2 and U20_2, respectively. These probability levels were selected to capture progressively rarer, larger-amplitude response peaks while retaining sufficient observations for meaningful comparison across simulation durations. The error bars indicate the 95% bootstrap confidence intervals. For U11_2, the 1-h, 3-h, and 5-h records give very similar results at $P_{\mathrm{ex}} = 0.10$, while the differences become more noticeable at lower exceedance probabilities, particularly for surge and fore mooring-line tension. Heave is comparatively insensitive to simulation duration, while pitch shows only moderate variation. The confidence intervals generally narrow with increasing simulation duration, most clearly at $P_{\mathrm{ex}} = 0.02$, indicating reduced sampling uncertainty for the longer records. A similar but weaker sensitivity of the peak amplitudes to simulation duration is observed for U20_2, where the three simulation durations give generally close peak amplitudes for all responses. The reduction in confidence-interval width from the 1-h to the longer records is again apparent, particularly at the lowest exceedance probability. The close agreement between the 3-h and 5-h results further suggests improved statistical stability of the peak estimates with increasing simulation duration.

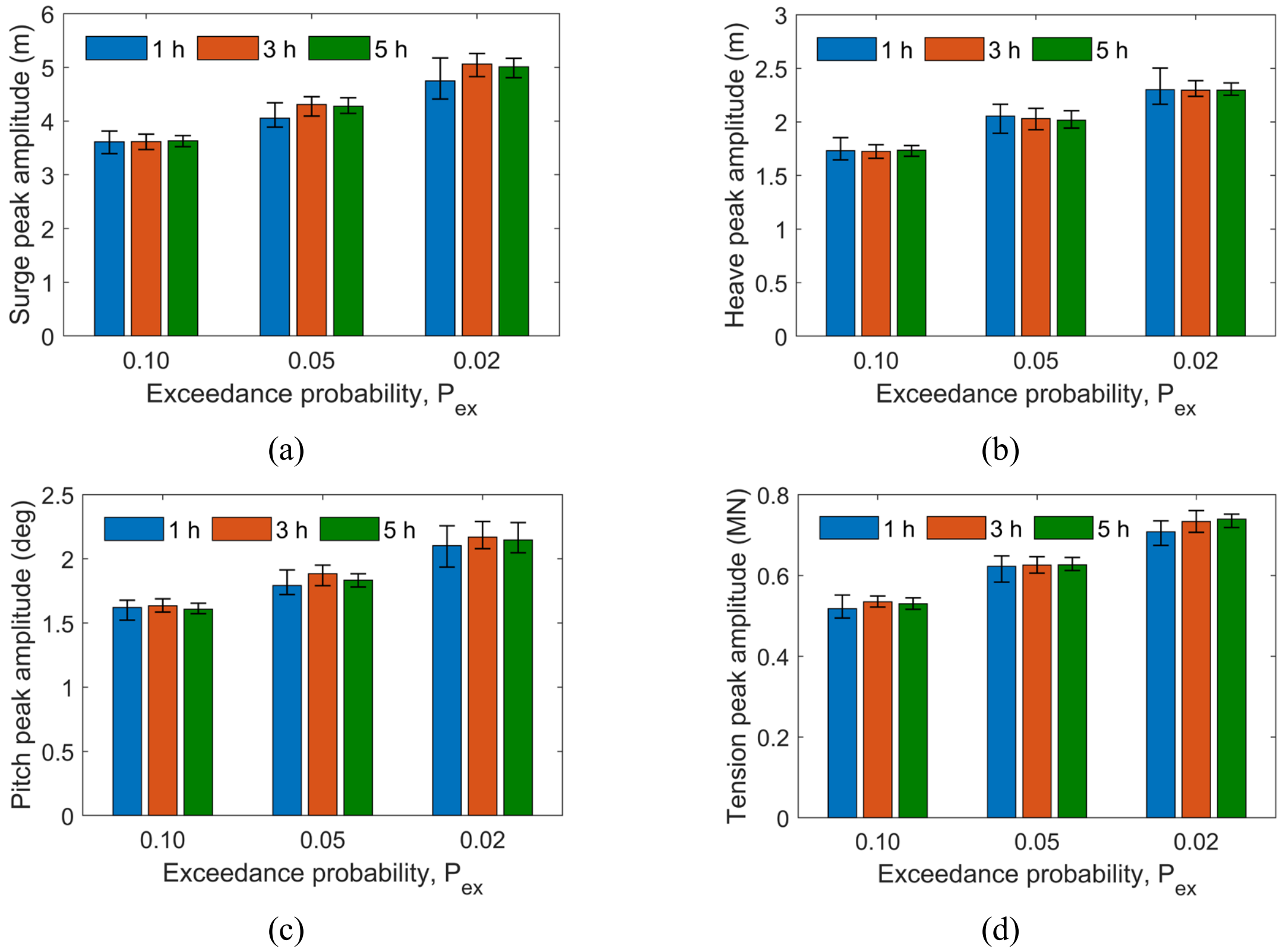

**Fig. 18.** Comparison of peak amplitudes at selected exceedance probabilities for different simulation durations in DLC1.6_U11_2. (a) Surge; (b) Heave; (c) Pitch; (d) Fore mooring-line tension. Error bars indicate the 95% bootstrap confidence intervals.

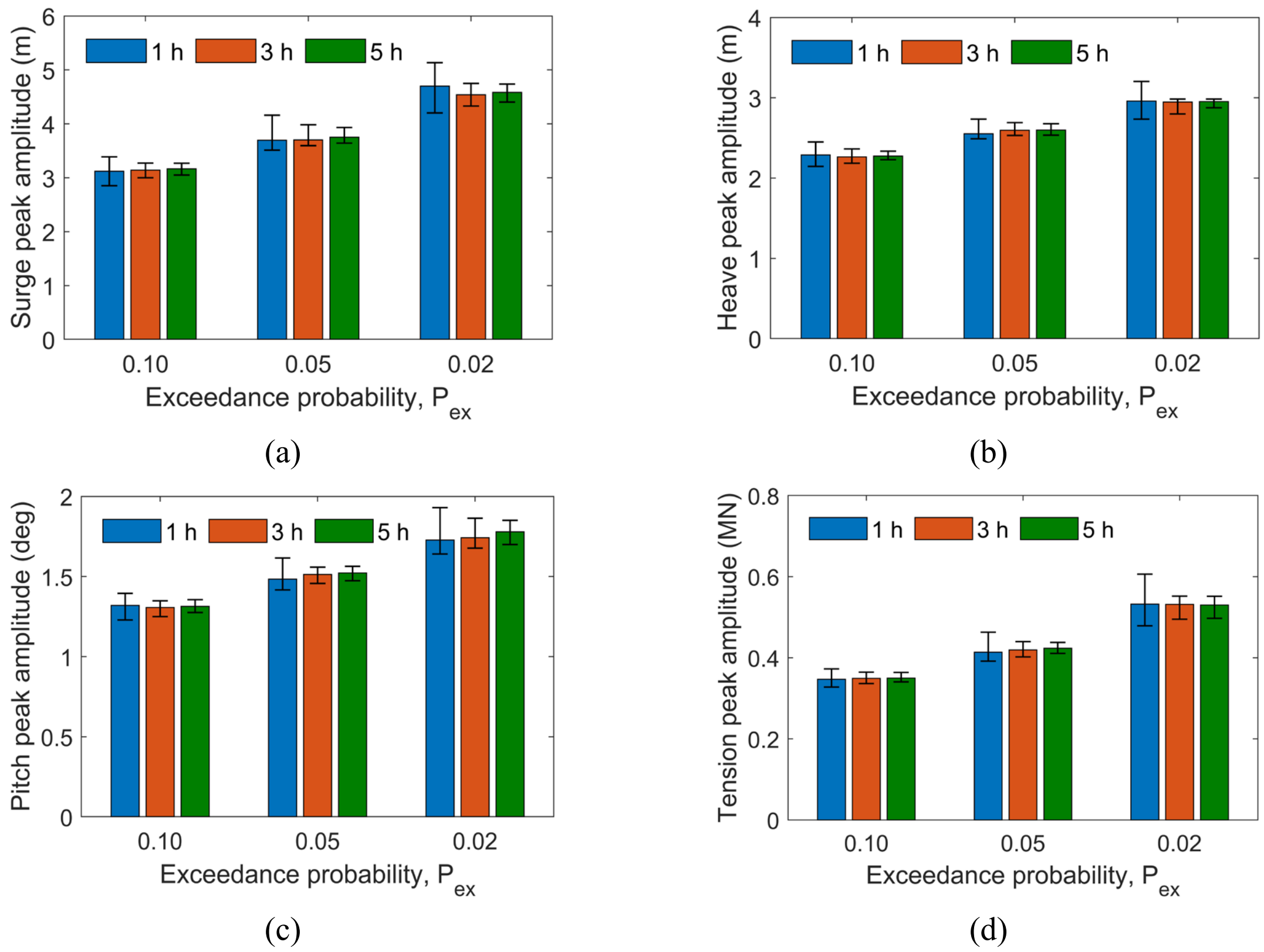

**Fig. 19.** Comparison of peak amplitudes at selected exceedance probabilities for different simulation durations in DLC1.6_U20_2. (a) Surge; (b) Heave; (c) Pitch; (d) Fore mooring-line tension. Error bars indicate the 95% bootstrap confidence intervals.

Overall, the 1-h records are adequate for characterising the bulk response statistics and spectral characteristics considered in the numerical–experimental comparisons, whereas the low-probability response tail is more sensitive to record length. For the present cases, extending the simulation from 1 h to 3 h noticeably improves the stability of the tail estimates, as also reflected by the narrower bootstrap confidence intervals. The peak-amplitude estimates and confidence intervals show comparatively small additional changes between 3 h and 5 h. These results indicate that the appropriate simulation duration depends on the response quantity of interest: relatively short records may be sufficient for estimating bulk response characteristics, whereas longer records are preferable for low-probability and extreme-response assessment, where sufficient sampling of rare, large-amplitude events is required to reduce statistical uncertainty and obtain more reliable estimates of the response tail.

## 5. Conclusions

This study investigated the role of second-order difference-frequency wave loading in the coupled response of a floating offshore wind turbine, with particular emphasis on when full-QTF modelling provides added value beyond linear hydrodynamics and Newman’s approximation. An efficient time-domain formulation using full QTFs with difference-frequency bandwidth truncation was incorporated into a fully coupled model of the IEA 15 MW turbine on the VolturnUS-S semi-submersible platform and compared with 1:70-scale real-time hybrid wave-basin experiments under wave-only and combined turbulent wind–wave conditions. The assessment covered response statistics, spectral characteristics, and peak-amplitude distributions, capturing both overall and low-frequency response differences among the hydrodynamic formulations.

Under wave-only conditions, difference-frequency excitation is important for capturing low-frequency surge and the associated mooring response, particularly in energetic seas when turbines are parked. Both Newman’s approximation and the Diff-QTF model substantially improve predictions over linear hydrodynamics, with the Diff-QTF model generally giving the closest agreement in low-frequency spectra and peak-amplitude distributions Under combined wind–wave conditions, the influence of difference-frequency loading cannot be inferred from sea-state severity alone. Strong low-frequency rotor thrust fluctuations dominate the surge and pitch responses, particularly near the rated wind speed, and substantially affect the mooring response through platform–mooring coupling. These responses therefore show reduced sensitivity to the hydrodynamic formulation. As these fluctuations weaken at higher wind speeds, the relative contribution of difference-frequency wave excitation becomes more pronounced, particularly for the low-frequency surge and mooring-tension responses. The Diff-QTF model generally provides the most consistent improvement in low-frequency and larger-amplitude surge and mooring responses, but its additional benefit over Newman’s approximation remains case-dependent.

The simulation-duration sensitivity analysis also confirmed that the 1-h records were adequate for characterising the bulk response statistics and spectral characteristics considered in the numerical–experimental comparison. The remaining underprediction under severe wave conditions, particularly in low-frequency surge and mooring responses, indicates that full-QTF-based difference-frequency loading alone is insufficient to reproduce all aspects of the measured low-frequency response. Nevertheless, across the more energetic wave cases, the predicted surge standard deviations remain within about 18% of the measurements, with smaller relative errors for heave, pitch, and mooring tension. Nonlinear hydrodynamic effects beyond the present formulation, including wave-induced viscous nonlinearities and strongly nonlinear free-surface effects, may contribute to this discrepancy and warrant further investigation. The present conclusions are limited to the investigated semi-submersible platform and test conditions. Future work should extend validation to a wider range of platform configurations and operating conditions. The current difference-frequency formulation is also being extended to multidirectional wave conditions by incorporating QTFs that account for both frequency and directional interactions.

## Acknowledgements

The authors gratefully acknowledge the financial support of EPSRC project FOWT-Control (Grant Nos. EP/W009854/1 and EP/W009692/1). Dr Yi Zhang is also supported by the Open Fund of State Key Laboratory of Coastal and Offshore Engineering in China with grant No. LP2502.

## Appendix A. Effect of difference-frequency force bandwidth

To improve computational efficiency while retaining the dominant low-frequency responses, the bandwidth of the difference-frequency force is truncated. Fig. A1 shows the effect of the difference-frequency force bandwidth $\Delta f$ specified in Eq. (10) on the PSDs of platform motions and fore mooring-line tension. The shorter-period wave-only case (Wave 2) is selected because its wave-frequency response is more clearly separated from the low-frequency range associated with the difference-frequency force, allowing the bandwidth effect to be more readily identified. A time step of 0.05 s is adopted throughout the analysis. Table A1 compares the standard deviations of surge, heave, pitch, and fore mooring-line tension, as well as the computational time for different $\Delta f$. It shows that the difference-frequency force bandwidth has a limited effect on the surge and mooring-tension PSDs but significantly influences the heave and pitch responses. In addition to the wave-frequency components, prominent response peaks occur near the natural frequencies of surge (0.008 Hz), heave (0.05 Hz), and pitch (0.033 Hz). Because the surge natural frequency is below all examined bandwidths, its resonant response is consistently retained. In contrast, $\Delta f$ = 0.025 Hz excludes excitation near the heave and pitch natural frequencies, underestimating their standard deviations by 15.6% and 26.1%, respectively. At $\Delta f$ = 0.05 Hz, the pitch resonance is captured, whereas the heave resonance is only partially covered, resulting in a 3.4% error. A bandwidth of $\Delta f$ = 0.075 Hz adequately captures all relevant resonant frequencies, limits the response errors to 0.38%, and reduces the computational time by around 75% relative to the full-band calculation. It therefore provides the best balance between accuracy and efficiency for Wave 2, and similar findings are observed for the other cases.

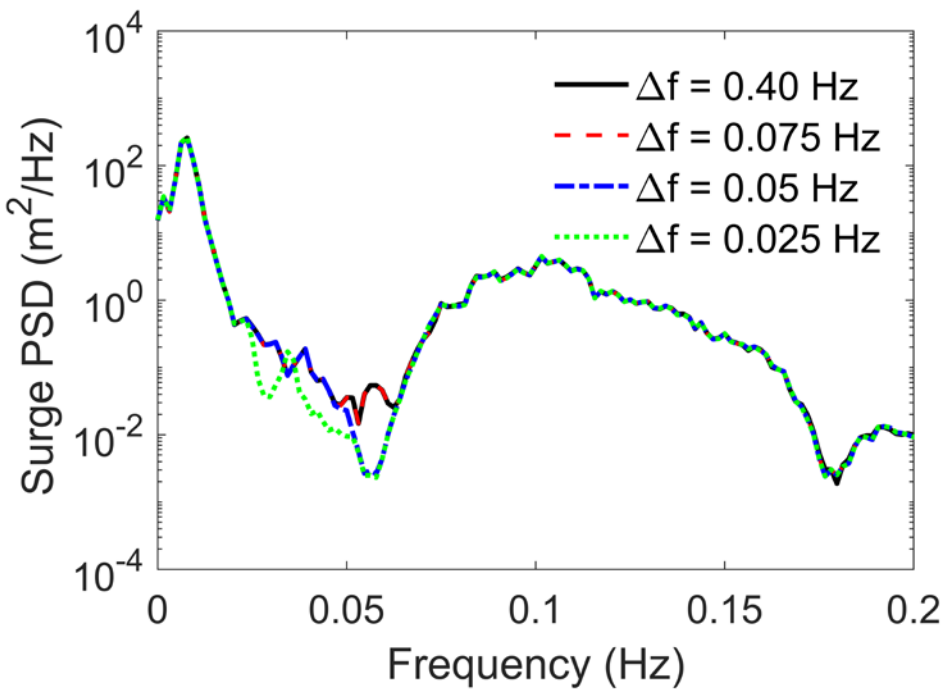

(a)

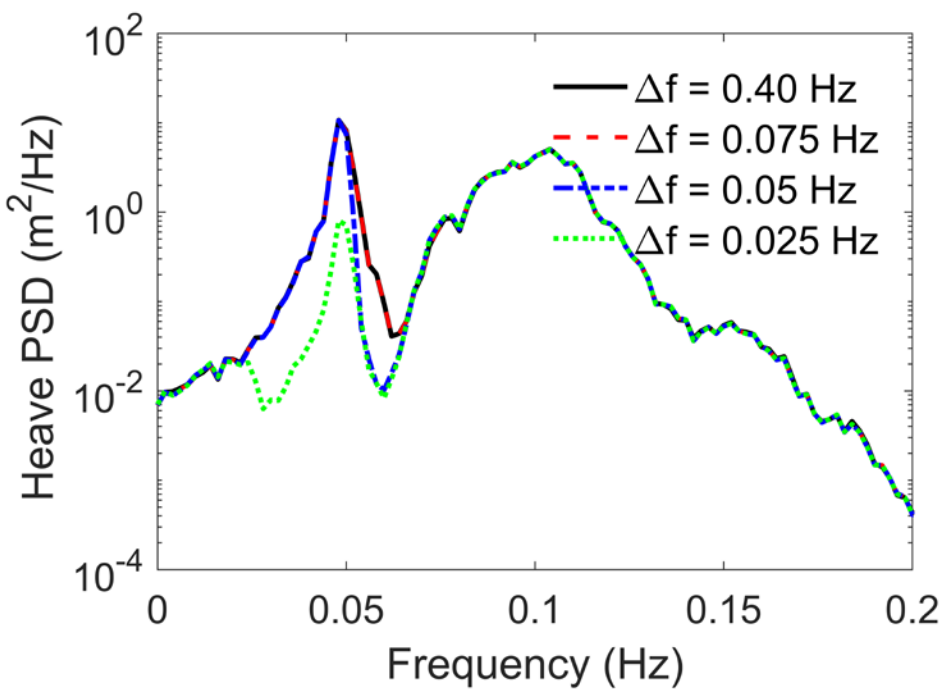

(b)

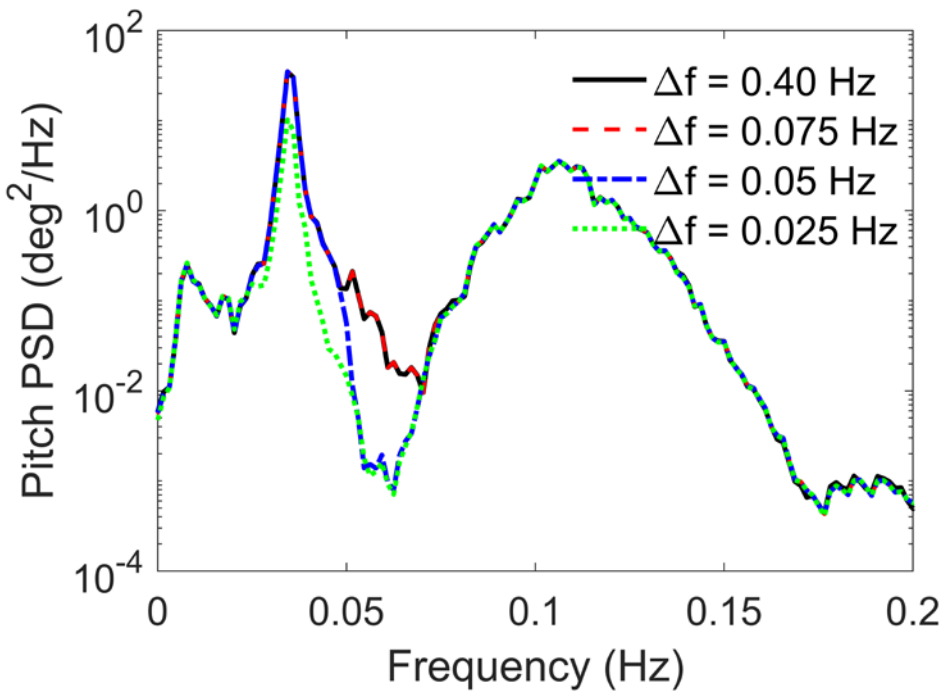

(c)

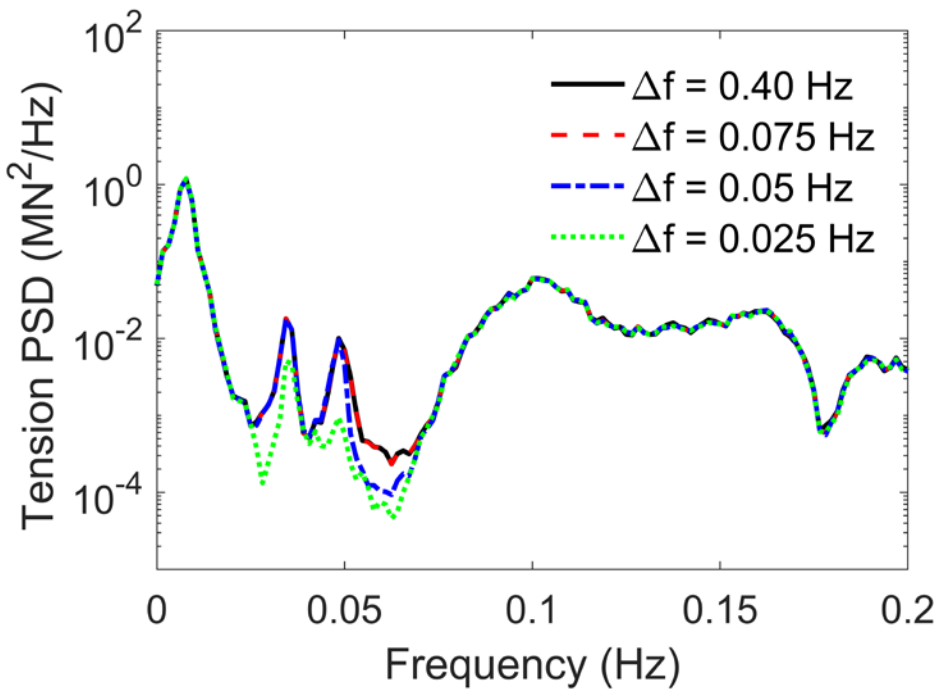

(d)

**Fig. A1.** Comparison of platform motion and mooring tension PSDs for different difference-frequency force bandwidth in Wave 2. (a) Surge PSD; (b) Heave PSD; (c) Pitch PSD; (d) Fore mooring-line tension PSD.

**Table A1.** Effects of difference-frequency force bandwidth on the simulation results and computational time for Wave 2. The values in parentheses are relative errors with respect to $\Delta f$ = 0.40 Hz (full band).

| $\Delta f$ (Hz) | Surge std (m) | Heave std (m) | Pitch std (deg) | Mooring tension std (MN) | CPU time (min) |
|---|---|---|---|---|---|
| 0.025 | 1.2717 (+0.14%) | 0.3704 (−15.6%) | 0.3524 (−26.1%) | 0.0904 (−0.33%) | 8.8 (−83.9%) |
| 0.05 | 1.2656 (−0.34%) | 0.4236 (−3.4%) | 0.4750 (−0.42%) | 0.0904 (−0.33%) | 10.9 (−80.1%) |
| 0.075 | 1.2711 (+0.09%) | 0.4383 (−0.09%) | 0.4788 (+0.38%) | 0.0908 (+0.11%) | 13.8 (−74.8%) |
| 0.40 (full band) | 1.2699 | 0.4387 | 0.4770 | 0.0907 | 54.7 |

## Appendix B. Time-step convergence test

Fig. B1 compares the time series of platform motions and fore mooring-line tension for different time steps ($\Delta t$) in DLC1.6_U11_1 (see Table 4), a near-rated wind and short-wave case with relatively high time-step sensitivity. The time histories show generally good agreement across different time steps, particularly for heave and pitch, while surge and mooring tension are more sensitive to the time step. This is mainly attributed to its low-frequency, lightly damped response, which is strongly influenced by the difference-frequency excitation near the surge natural frequency. Although the difference-frequency force itself remains essentially unchanged with time step, a coarser time step can introduce additional numerical damping and phase error in the dynamic integration, thereby suppressing the surge resonance. Table B1 compares the corresponding standard deviations and maximum values of the responses for different $\Delta t$. As $\Delta t$ decreases from 0.10 to 0.025 s, the standard deviations of surge and fore mooring-line tension increase by about 8.1% and 7.2%, respectively, whereas their maxima vary by only about 1% due to their large mean components. The differences between $\Delta t$ = 0.05 and 0.025 s are considerably smaller, with differences in the standard deviations below 2% and changes in the maximum responses below 0.4% for all quantities. Considering that the computational time for this case nearly doubles from 32.1 to 63.2 minutes, $\Delta t$ = 0.05 s provides a reasonable compromise between numerical accuracy and computational efficiency and was therefore adopted for all cases.

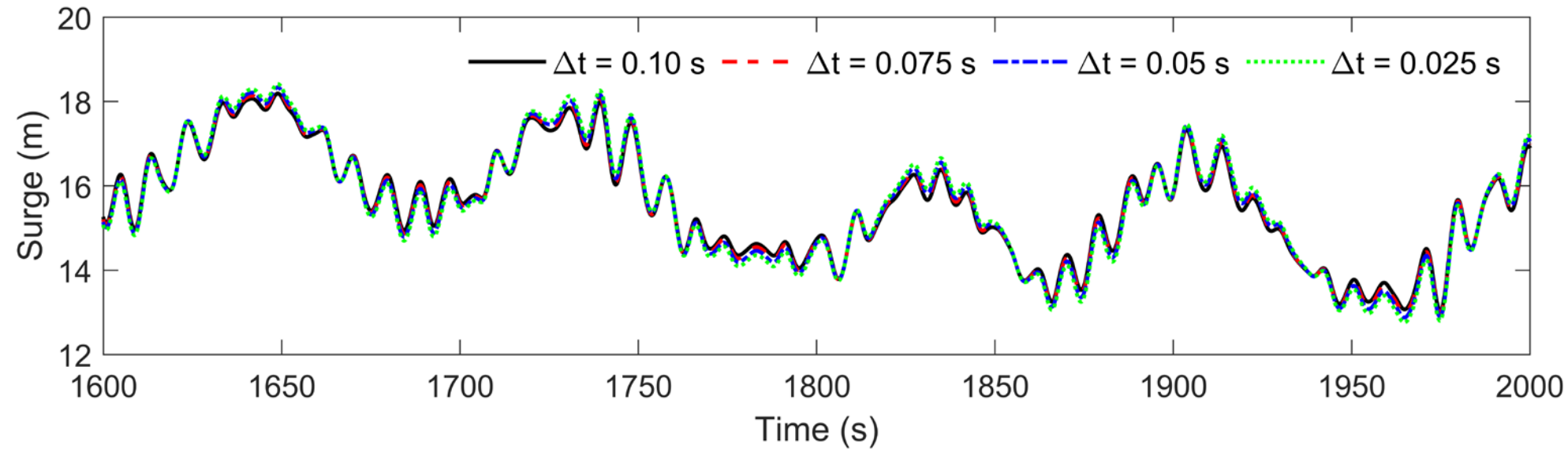

(a)

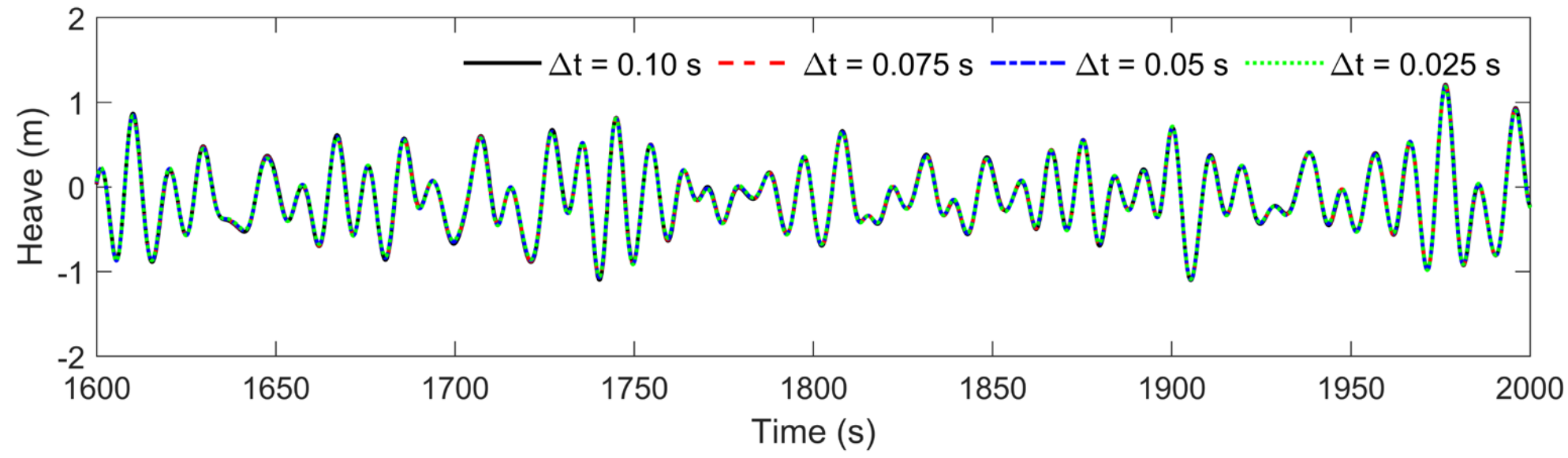

(b)

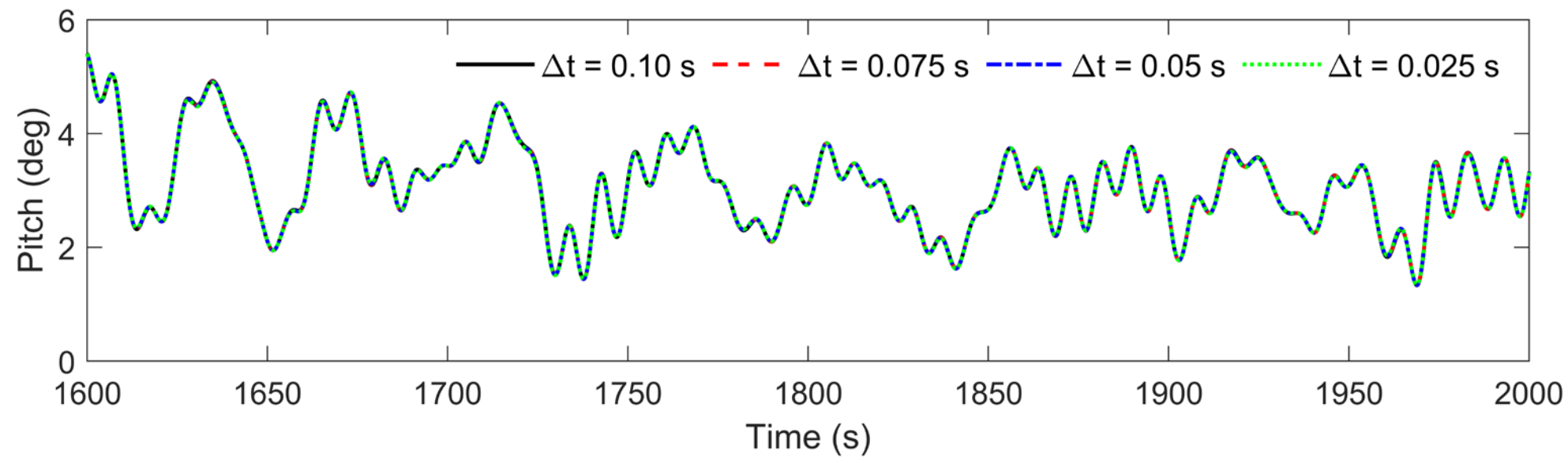

(c)

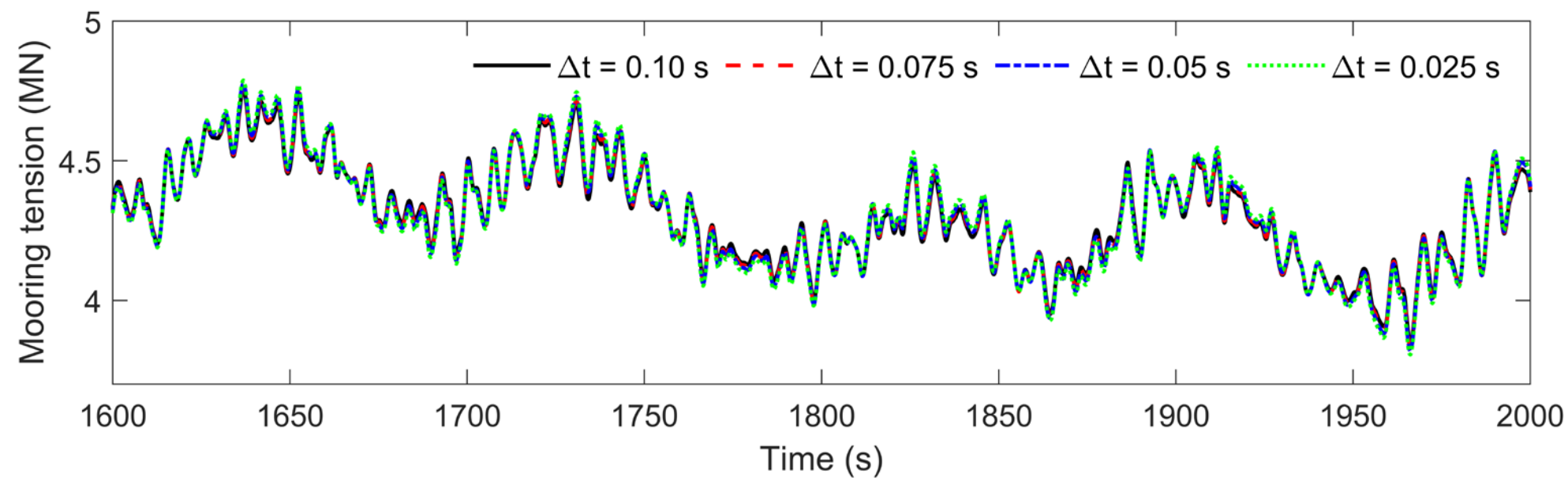

(d)

**Fig. B1.** Comparison of the time series of platform motions and mooring tension obtained using different time steps for DLC1.6_U11_1. (a) Surge; (b) Heave; (c) Pitch; (d) Fore mooring-line tension.

**Table B1.** Time-step sensitivity of the simulation results for DLC1.6_U11_1.

| $\Delta t$ (s) | Surge std / max (m) | Heave std / max (m) | Pitch std / max (deg) | Mooring tension std / max (MN) | CPU time (min) |
|---|---|---|---|---|---|
| 0.10 | 2.0407 / 22.203 | 0.4522 / 1.4915 | 0.8949 / 6.1048 | 0.2924 / 5.2855 | 19.4 |
| 0.075 | 2.0997 / 22.312 | 0.4480 / 1.4878 | 0.8944 / 6.1144 | 0.3000 / 5.3021 | 26.3 |
| 0.05 | 2.1647 / 22.426 | 0.4440 / 1.4840 | 0.8940 / 6.1259 | 0.3084 / 5.3183 | 32.1 |
| 0.025 | 2.2067 / 22.509 | 0.4412 / 1.4795 | 0.8936 / 6.1380 | 0.3135 / 5.3284 | 63.2 |